\documentclass[lettersize,journal]{IEEEtran}

\usepackage{amsmath,amssymb,amsfonts,bm}
\usepackage{xcolor}
\usepackage{array}
\usepackage{textcomp}
\usepackage{stfloats}
\usepackage{url}
\usepackage{verbatim}
\usepackage{graphicx}
\usepackage{pifont}
\usepackage{cite}
\usepackage{subfigure}
\usepackage{hyperref}
\begin{document}

\title{Active Sensing for Multiuser Beam Tracking with Reconfigurable Intelligent Surface}

\author{Han Han,~\IEEEmembership{Member,~IEEE,} Tao Jiang,~\IEEEmembership{Member,~IEEE,} and Wei Yu,~\IEEEmembership{Fellow,~IEEE} 
\thanks{Manuscript accepted in IEEE Transactions on Wireless Communications. This work was supported by Huawei Technologies Canada. The materials in this paper have been presented in part at the IEEE International Conference on Acoustics, Speech and Signal Processing (ICASSP), 2023 \cite{10097134}. The source code for the numerical results in this paper is available at \href{https://github.com/HanHanCoding/MUTracking-ActiveSensing}{https://github.com/HanHanCoding/MUTracking-ActiveSensing}. (\textit{Corresponding author: Wei Yu.})}
\thanks{Han Han was with the Edward S. Rogers Sr. Department of Electrical and Computer Engineering, University of Toronto, Toronto, ON M5S 3G4, Canada. He is now with Marvell Semiconductor Canada, Inc., Toronto, ON, Canada. (e-mail: johnny.han@mail.utoronto.ca).}
\thanks{Tao Jiang was with the Edward S. Rogers Sr. Department of Electrical and Computer Engineering, University of Toronto, Toronto, ON M5S 3G4, Canada. He is now with Qualcomm Technologies, Inc., Santa Clara, CA, USA. (e-mail: taoca.jiang@mail.utoronto.ca).}
\thanks{Wei Yu is with The Edward S. Rogers Sr. Department of Electrical and Computer Engineering, University of Toronto, Toronto, ON M5S 3G4, Canada. (e-mail: weiyu@ece.utoronto.ca).}
}

\maketitle
\thispagestyle{empty}

\begin{abstract}
This paper studies a beam tracking problem in which an access point (AP), in
collaboration with a reconfigurable intelligent surface (RIS), dynamically adjusts its downlink beamformers and the reflection pattern at the RIS in order to maintain reliable communications with multiple mobile user equipments (UEs).  
Specifically, the mobile UEs send uplink pilots to the AP periodically during
the channel sensing intervals, the AP then adaptively configures the beamformers 
and the RIS reflection coefficients for subsequent data transmission based on 
the received pilots.  This is an active sensing problem, because channel
sensing involves configuring the RIS coefficients during the pilot stage and
the optimal sensing strategy should exploit the trajectory of channel state
information (CSI) from previously received pilots. Analytical solution to such an active
sensing problem is very challenging. In this paper, we propose a deep learning
framework utilizing a recurrent neural network (RNN) to automatically summarize
the time-varying CSI obtained from the periodically received pilots into state
vectors. These state vectors are then mapped to the AP
beamformers and RIS reflection coefficients for subsequent downlink data
transmissions, as well as the RIS reflection coefficients for the next round
of uplink channel sensing. The mappings from the state vectors to the downlink
beamformers and the RIS reflection coefficients for both channel sensing and
downlink data transmission are performed using graph neural networks (GNNs) to
account for the interference among the UEs. Simulations demonstrate
significant and interpretable performance improvement of the proposed approach
over the existing data-driven methods with nonadaptive channel sensing schemes.  
\end{abstract}


\section{Introduction}

Reconfigurable intelligent surface (RIS) is a promising technology for future
wireless communication systems due to its ability to reflect incoming signals 
toward desired directions in an adaptive fashion \cite{9140329}. The RIS can enhance
communications by establishing focused beams between the access point (AP) and
the user equipment (UE), but this focusing capability depends crucially on the
availability of channel state information (CSI), which must be obtained in a dedicated 
channel sensing stage using pilot signals \cite{9087848,9366805,9130088}.
However, in high-mobility scenarios where CSI needs to be measured frequently, 
estimating CSI from scratch in each channel sensing phase can lead to a large
pilot overhead. The main idea of this paper is that significant saving in pilot
overhead is possible by exploiting the temporal channel correlation due to UE mobility in the channel sensing stage.

This paper considers a multiuser beam tracking problem in an RIS-assisted
mobile communication system, in which the mobile UEs periodically send pilots
to the AP through reflection at an RIS, and the AP designs the downlink RIS
reflection coefficients and the beamforming vectors based on the received
pilots to maintain beam alignment with the mobile UEs over time. To alleviate
pilot training overhead, this paper considers the incorporation of
\emph{active sensing} strategy \cite{sohrabi2022active} into the beam tracking
process. In the proposed active sensing scheme, the RIS reflection
coefficients in the CSI acquisition stages are adaptively designed based on the
pilots received previously. In effect, we design the RIS to keep track of the
mobile UEs with focused sensing beams during the pilot stage. Subsequently, the downlink RIS reflection coefficients and AP beamformers are also designed to focus precisely towards the mobile UEs for data transmission.

The multiuser beam tracking problem in an RIS-assisted mobile communication system is highly nontrivial, primarily due to the significant pilot overhead required to maintain beam alignment. This is because RIS cannot perform active signal transmission or reception. Consequently, the high-dimensional channels to and from the RIS can only be estimated indirectly based on the low-dimensional observations at the AP. Utilizing active sensing to leverage historical channel information can significantly reduce pilot overhead; but finding the optimal active sensing strategy is very challenging due to the complexity of optimization over a sequence of observations and actions.

To address these challenges, this paper proposes a novel deep active sensing framework for the multiuser beam tracking problem. We employ recurrent neural networks (RNNs) with long short-term memory (LSTM) units \cite{6795963} to automatically summarize information from the historical observations into state vectors. These state vectors are used to design the RIS reflection coefficients for both channel sensing and data transmission stages, and to design the downlink AP beamformers. To efficiently manage interference among the UEs, we utilize a graph neural network (GNN) \cite{hamilton2017inductive} to model the spatial relation between the UEs and the RIS, and exploit this information to optimize the mappings from the state vectors to the beamformers and reflection patterns. The proposed framework can be shown to significantly reduce pilot overhead, while producing interpretable beamforming and reflection patterns.

\subsection{Related Works}
The channel tracking problems in RIS-assisted mobile communication systems have been investigated previously using model-based statistical signal processing tools, such as Kalman filtering and its variants \cite{zegrar2020general,10250189,9772982,cai2021downlink,9854102}. For example, \cite{zegrar2020general} studies channel tracking in an RIS-assisted millimeter wave (mmWave) system by using a linear state-space equation to model the temporal correlations of the time-varying RIS-UE channel, which is parameterized by the fading coefficients and the angles of arrival/departure. The temporal channel correlations can then be exploited to reduce pilot overhead using an extended Kalman filter. Additional studies \cite{10250189, 9772982, cai2021downlink, 9854102} further apply Kalman filtering into various beam tracking contexts, such as the multiple-input multiple-output (MIMO) systems aided by two RISs \cite{9772982} and the RIS-assisted frequency division duplexing (FDD) systems \cite{cai2021downlink}. While Kalman filtering based algorithms offer low computational complexity, they require explicit modeling of the temporal channel correlations by using a state transition model \cite{va2016beam}. In practical mobile communication scenarios where temporal channel correlations is not easily captured by models, the mismatch between the time-varying environment and the state transition models can significantly degrade performance.

To address the issue of model mismatch, studies including
\cite{10109153,9869783,9743298,liu2022scalable} propose using machine learning
algorithms to directly learn the temporal channel correlations in a data-driven
manner for RIS-assisted beam tracking. Temporal correlation can be captured in
a data-driven way either using RNN or by reinforcement learning.  For example,
\cite{10109153,9869783} employ reinforcement learning to capture the temporal
correlations in RIS-assisted unmanned aerial vehicle (UAV) systems by
formulating the beam tracking problem as a Markov decision process (MDP). 
The action space includes the design of RIS reflection coefficients and
beamformers, the state space comprises prior actions along with resulting UAV
coordinates and corresponding CSI, and the reward function is defined as the system utility. 
However, reinforcement learning is
known to suffer from slow convergence \cite{dulac2021challenges}. Moreover, a
correct definition of state is crucial in MDP, but it is not easy to define
state properly in many problems. For example, \cite{10109153,9869783} use perfect CSI as part of the MDP states, but perfect CSI is rarely available in practice. Instead of
reinforcement learning, \cite{9743298} proposes to learn the system state for
channel tracking in RIS-aided UAV systems using an RNN. Likewise, RNN is used
in \cite{liu2022scalable} for the RIS-assisted multiuser beam tracking problem.
The RNN has an ability to automatically summarize time-varying CSI based on the received pilots into a learned state vector. This is a
crucial advantage that motivates us also to use the RNN in our setting.

The problem setting in this paper also requires the modeling of spatial
relationship between the AP, the RIS, and the UEs.
Toward this end, GNN is an indispensable tool for learning these spatial relations
\cite{9072356, 9285223, 9252917,9844981}.
Specifically, previous studies including
\cite{jiang2021learning,9072356,9844981,liu2022scalable} propose using GNNs to
efficiently manage the mutual interference among the UEs by matching the graph
structure of the neural network to the spatial connections between the UEs and
the RIS. This is an important architectural feature that we also adopt in this paper. 
It allows more efficient training and gives more interpretable results.

One of the key differentiating features of this paper as compared to prior works
is in the design of \emph{active sensing} strategies for channel sensing. By active sensing, we mean that the AP adaptively and sequentially queries the environment to measure and to track the channel dynamics, by using a sequence of focused sensing beams towards the mobile UEs. In particular, at each CSI acquisition
stage, the AP adaptively designs the sensing patterns at the RIS based on the information obtained from all the previous stages. In most prior works
\cite{zegrar2020general,10250189,9772982,cai2021downlink,9854102,10109153,9869783,9743298},
the CSI acquisition stage uses fixed RIS reflection coefficients, either generated randomly or based on fixed patterns such as the discrete Fourier transform (DFT) \cite{9053695}. 
However, these non-adaptive RIS sensing schemes are far from optimal. On the other
hand, \cite{liu2022scalable} adaptively designs the RIS sensing scheme based on
an RNN, but proposes to employ the same RIS reflection coefficients in both
sensing and data transmission stages. This is suboptimal due to the
different functionalities of the RIS in two stages.  A main contribution of
this paper is a learning approach for adaptively designing the 
RIS coefficients in the channel sensing stage for tracking multiple UEs, which
to the best of authors' knowledge has not been explored in the existing literature. 
Recently, \emph{active sensing} has demonstrated remarkable performance in related 
applications including localization \cite{10373816} and beam alignment \cite{8792366,9448070,sohrabi2022active,10124207,10304519} in fixed channel environment, where the goal is to
leverage previously obtained channel measurements to gradually refine focusing onto some desired
low-dimensional part of the static channel. The current investigation is concerned with 
active sensing for beam tracking in the time-varying channels.

\subsection{Main Contributions}
\subsubsection{Active Sensing Using RNN}
This paper proposes to leverage historical channel information to adaptively
design RIS reflection coefficients across multiple pilot stages to address a multiuser
beam tracking problem. This is accomplished by adopting an LSTM-based RNN for
its ability to capture the temporal correlations in the channel dynamics. At
each CSI acquisition stage, a set of LSTM cells receive a new round of pilots
from the mobile UEs, then update its fixed-dimensional hidden and cell state
vectors accordingly.  The cell state vectors are subsequently utilized to
design the RIS reflection coefficients for the next round of channel sensing
(as well as for data transmission).  The proposed active sensing strategy can
achieve significant performance gain over the benchmarks with non-adaptive
sensing schemes. The proposed design also exhibits robust performance beyond the training range as it is able to update channel information based on the latest received pilots.

\subsubsection{Reflection Coefficients Optimization Using GNN}
This paper proposes to use GNN to learn the mappings from the LSTM cell state
vectors to the RIS reflection coefficients for both channel sensing and data
transmission. The GNN leverages the spatial relation
between the UEs and the RIS to design the reflection patterns while accounting for
the mutual interference between the UEs.
The proposed GNN architecture
extends the designs in \cite{liu2022scalable,jiang2021learning,9783100},
while specifically addressing the distinct roles of the RIS reflection patterns in the sensing 
vs. data transmission stages. The proposed GNN improves upon previous work
\cite{liu2022scalable}
that uses the same RIS coefficients for both sensing and communication. 

\subsubsection{Downlink AP Beamforming}
After each CSI acquisition stage, this paper proposes to estimate the
low-dimensional effective channels between the AP and the UEs by transmitting
additional short pilots. 
The AP beamformers can then be analytically designed. 
This allows additional performance gain with only a few additional pilots.


\subsection{Organization of the Paper and Notations}
The remaining parts of this paper are organized as follows. Section \ref{System Model and Problem Formulation} introduces the system model and pilot transmission protocol. Section \ref{Active Sensing for Tracking} presents the proposed active sensing scheme for beam tracking. Section \ref{Proposed Data-Driven Approach} introduces the proposed learning-based active sensing framework. Numerical results and visual interpretations are provided in Section \ref{Numerical Results and Interpretations}. Section \ref{conclusion} concludes the paper.

We use lowercase letters, lowercase bold-faced letters, and uppercase bold-faced letters to denote scalars, vectors, and matrices, respectively. We use $(\cdot)^{\sf{T}}$ and $(\cdot)^{\sf{H}}$ to denote transpose and Hermitian transpose, respectively. We use $\mathcal{CN}(\cdot,\cdot)$ and $\mathcal{U}(\cdot,\cdot)$ to denote complex Gaussian distributions and continuous uniform distributions, respectively. 

\section{System Model and Problem Formulation}\label{System Model and Problem Formulation}
\subsection{System Model}
\begin{figure}[t]
	\includegraphics[scale=0.55]{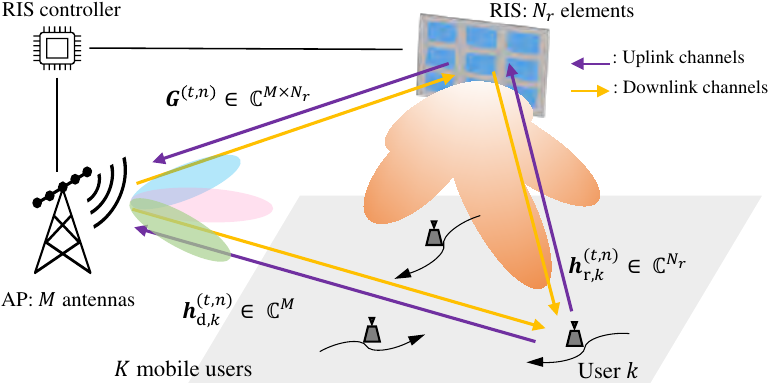}
	\centering
    \caption{RIS assisted multiuser mobile communication system.}
	\label{fig:Channel Model}
\end{figure}
Consider a narrowband RIS-assisted multiuser multiple-input single-output
(MU-MISO) system where $K$ single-antenna user equipment (UEs) are served by 
an access point (AP) with $M$ antennas. The RIS, equipped with $N_r$ passive
elements, is placed between the AP and UEs to enable the reflection links as
shown in Fig.~\ref{fig:Channel Model}. The RIS is assumed to operate under ideal conditions with perfect phase control and negligible hardware impairments. The RIS controller works cooperatively with the AP to control the magnitudes and directions of the reflected signals by tuning the phase shifters of the passive elements of the RIS. We model the time-varying channel in the discrete-time domain, where the continuous evolution of the channel is approximated by dividing time into a sequence of fixed-length blocks. The channel coefficients are constant within the block and are correlated across the successive blocks. This correlation reflects the gradual changes in the channel due to the mobility of the UEs \cite{938579,995514}. We further define a framing structure 
consisting of $N+1$ blocks per frame, and assume that channel variation is not 
significant within the frame so a fixed beamforming strategy across all the blocks within the frame can be used 
without incurring substantial performance loss.  
For the $n$-th block of the $t$-th frame, we use $\bm{G}^{(t,n)} \in
\mathbb{C}^{M \times N_r}$, $\bm{h}_{\mathrm{d},k}^{(t,n)} \in \mathbb{C}^{M}$,
and $\bm{h}_{\mathrm{r},k}^{(t,n)} \in \mathbb{C}^{N_r}$ to denote the channel
matrix from the RIS to the AP, the channel vector from the UE $k$ to the AP,
and the channel vector from the UE $k$ to the RIS, respectively.

In the $n$-th block of the $t$-th frame, let $s_{k,i}^{(t,n)} \in \mathbb{C}$ be the $i$-th data symbol transmitted from the AP to the UE $k$, with $\mathbb{E}[|s_{k,i}^{(t,n)}|^2]=1$. The AP transmits $s_{k,i}^{(t,n)}$ to the UE $k$ through a beamforming vector $\bm{b}_k^{(t)} \in \mathbb{C}^{M}$, subject to a power constraint $\sum_{k=1}^K\|\bm{b}_k^{(t)}\|_2^2 \leq P_d$, where $P_d$ is the downlink transmit power at the AP. We use $\bm{B}^{(t)} = [\bm{b}_1^{(t)}, \dots, \bm{b}_K^{(t)}] \in \mathbb{C}^{M \times K}$ to denote the beamforming matrix at the AP in the $t$-th frame. Let $\bm{w}^{(t)} = [e^{j\theta_{1,t}}, e^{j\theta_{2,t}}, \dots, e^{j\theta_{N_r,t}} ]^{\sf{T}} \in \mathbb{C}^{N_r}$ be the downlink RIS reflection coefficients in the $t$-th frame, where $\theta_{u,t} \in [0, 2\pi)$ is the phase shift of the $u$-th passive RIS element. Then, the signal $y_{k,i}^{(t,n)} \in \mathbb{C}$ received at the UE $k$ in the $n$-th block of the $t$-th frame can be expressed as:
\begin{equation}
    y_{k,i}^{(t,n)} = \sum_{j=1}^K (\bm{A}_k^{(t,n)} \widetilde{\bm{w}}^{(t)} )^{\sf{H}} \bm{b}_j^{(t)} s_{j,i}^{(t,n)} + z_{k,i}^{(t,n)} , \label{eq:DL signal}
\end{equation}
where $\bm{A}_k^{(t,n)} \triangleq [\bm{h}_{\mathrm{d},k}^{(t,n)}, \bm{G}^{(t,n)} \operatorname{diag}(\bm{h}_{\mathrm{r},k}^{(t,n)})] \in \mathbb{C}^{M \times (N_r + 1)}$ is the combined channel between the AP and the UE $k$ in the $n$-th block of the $t$-th frame, $\widetilde{\bm{w}}^{(t)} \triangleq [1, (\bm{w}^{(t)})^{\sf{T}}]^{\sf{T}} \in \mathbb{C}^{N_r + 1}$ is the effective RIS reflection coefficients, and $z_{k,i}^{(t,n)} \sim \mathcal{CN}(0, \sigma^2_d)$ denotes the i.i.d. Gaussian noise at the UE $k$. Therefore, the downlink achievable data rate $R_k^{(t,n)}$ for the UE $k$ in the $n$-th block of the $t$-th transmission frame can be expressed as:
\begin{equation}
R_k^{(t,n)} \!= \log \left(\! 1 \!+\! \frac{|(\bm{A}_k^{(t,n)} \widetilde{\bm{w}}^{(t)})^{\sf{H}} \bm{b}_k^{(t)}|^2}{\sum_{j=1, j \neq k}^K|(\bm{A}_k^{(t,n)} \widetilde{\bm{w}}^{(t)})^{\sf{H}} \bm{b}_j^{(t)}|^2 \!+\! \sigma_d^2}\right) .
\end{equation}

In this paper, our goal is to jointly optimize the beamforming matrix $\bm{B}^{(t)}$ and the RIS reflection coefficients $\bm{w}^{(t)}$ to establish $K$ reliable communication links between the AP and $K$ mobile UEs as well as to ensure fairness across the UEs. Accordingly, the optimization problem is formulated on a per-frame basis as:
\begin{equation} \label{eq: optimization - 1}
	\begin{aligned}
		(\mathrm{P}1): \;\;\; \mathop{\mathrm{maximize}}_{ \bm{B}^{(t)}, \ \bm{w}^{(t)} } \;\;
		& \; \frac{1}{N}\sum_{n=1}^N \left(\min _k R_k^{(t,n)} \right) \\ 
		\mathop{\mathrm{subject\ to}} \;\;
		& {}_{\mathop{}}  
		\sum_{k=1}^K\|\bm{b}_k^{(t)}\|_2^2 \leq P_d  \\
        & {}_{\mathop{}}  
        \left\lvert[\bm{w}^{(t)}]_i\right\rvert = 1, \ \forall i,t.
	\end{aligned}
\end{equation}

The optimization problem $(\mathrm{P}1)$ is a challenging nonconvex program. Moreover, solving problem $(\mathrm{P}1)$ requires the knowledge of the combined channels $\bm{A}_k^{(t,n)}$, which must be estimated. This paper assumes the system operates in time division duplex (TDD) mode, so we can leverage uplink-downlink channel reciprocity to acquire the instantaneous CSI from uplink pilots for downlink beamforming and for setting the RIS reflection coefficients. 

\begin{figure}[t]
	\includegraphics[scale=0.72]{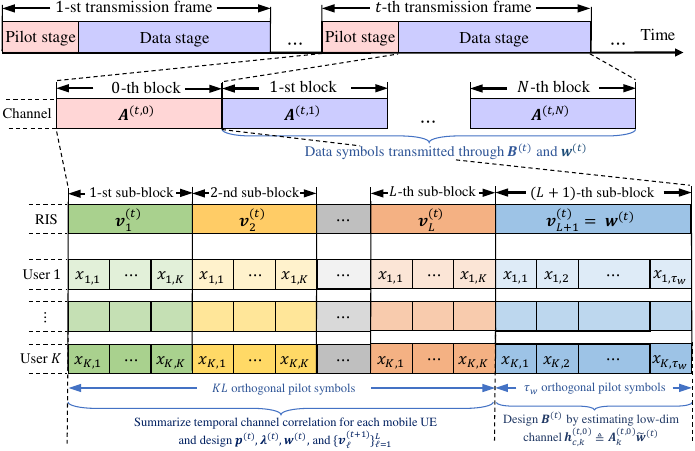}
	\centering
	\caption{Frame structure of the proposed transmission protocol.}
	\label{fig:Frame Structure}
\end{figure}

\subsection{Uplink Pilot Transmission Protocol}
The uplink pilot transmission protocol is shown in Fig.~\ref{fig:Frame Structure},
in which the mobile UEs send pilots to the AP in the $0$-th block of each
transmission frame (also referred to as the pilot stage.) Moreover, this $0$-th block is divided into $L+1$ sub-blocks. 
The first $L$ sub-blocks consists of $\tau=K$ symbols; and the last sub-block consists of
$\tau=\tau_w$ symbols, thus the pilot stage (i.e., the $0$-th block) has a total of $K L + \tau_w$ symbols.

Here, the pilots in the first $L$ sub-blocks serve to design the
RIS reflection coefficients, while the pilots in the last
sub-block are utilized for designing the AP beamformers. 
The pilot sequences $\bm{x}_k$ of the UEs are designed to be orthogonal to
each other so that they can be decorrelated at the AP, i.e.,
$\bm{x}_i^{\sf{H}}\bm{x}_j = 0$ if $i \neq j$ and $\bm{x}_i^{\sf{H}}\bm{x}_i =
\tau P_u$, where $P_u$ is the uplink pilot transmission power, $\tau$ is the
pilot length, and $i,j \in \{1,2,\dots,K\}$. The same pilots are sent repeatedly over
the first $L$ sub-blocks to design the downlink transmission RIS coefficients in the current 
frame (as well as the RIS sensing coefficients in the next frame.) In the last sub-block, 
we fix the downlink transmission RIS coefficients then send $\tau_w$ additional pilots to estimate the effective low-dimensional channel.

A crucial observation here is that the pilots are received through the RIS
reflection path. Thus, the RIS reflection coefficients are design parameters 
that can be set in each pilot stage. Here, we assume that the $L$ RIS reflection
coefficients across the $L$ sub-blocks are designed together within each frame and 
adaptively across the frames, so that the UE-to-RIS
and the RIS-to-AP channels are sensed in a sequential fashion in order to account for the historical channel observations.  We refer to the set of $L$ RIS reflection
coefficients in the pilot stage of each frame as the \textit{RIS sensing vectors}. 

In all $L+1$ sub-blocks, the AP decorrelates the received pilots in each sub-block by matching the pilot sequence for each UE. Let $\widehat{\bm{Y}}_{\ell}^{(t)} = [\widehat{\bm{y}}_{(\ell-1) \tau+1}^{(t)}, \dots, \widehat{\bm{y}}_{\ell \tau}^{(t)}] \in \mathbb{C}^{M\times \tau}$ denote the received signal at the AP in the $\ell$-th sub-block of the pilot stage in the $t$-th frame. 
 Let $\bm{v}^{(t)}_\ell \in \mathbb{C}^{N_r}$ denote the $\ell$-th sensing vector in the pilot stage of the $t$-th frame, we can express 
\begin{equation}\label{eq:pilots_1}
    \widehat{\bm{Y}}_{\ell}^{(t)} = \sum_{k=1}^K \bm{A}_k^{(t,0)} \widetilde{\bm{v}}^{(t)}_\ell \bm{x}_{k}^{\sf{H}} + \widehat{\bm{Z}}_{\ell}^{(t)} ,
\end{equation}
where $\widetilde{\bm{v}}^{(t)}_\ell \triangleq [1, (\bm{v}^{(t)}_\ell)^{\sf{T}}]^{\sf{T}} \in \mathbb{C}^{N_r + 1}$ is the $\ell$-th effective sensing vector in the $t$-th frame and $\widehat{\bm{Z}}_{\ell}^{(t)} \in \mathbb{C}^{M\times \tau}$ is the sensing noise, whose columns are assumed to be independently distributed as $\mathcal{CN}(0, \sigma^2_{u}\bm{I})$. By the orthogonality of the pilots, we can form $\bar{\bm{y}}_{k,\ell}^{(t)} \in \mathbb{C}^M$, which is the contribution from the UE $k$ in the $\ell$-th sub-block of the pilot stage within the $t$-th frame \cite{10053657,jiang2021learning}:
\begin{equation} \label{eq: pilots in L+1 block}
    \bar{\bm{y}}_{k,\ell}^{(t)} = \frac{1}{\tau P_u} \widehat{\bm{Y}}_{\ell}^{(t)} \bm{x}_{k} = \bm{A}_k^{(t,0)} \widetilde{\bm{v}}^{(t)}_\ell + \bar{\bm{z}}_{k, \ell}^{(t)} ,
\end{equation}
where $\bar{\bm{z}}_{k, \ell}^{(t)} \! \triangleq \! \frac{1}{\tau P_u} \widehat{\bm{Z}}_{\ell}^{(t)} \bm{x}_{k}$. Denote $\bar{\bm{Y}}_{k}^{(t)} = [\bar{\bm{y}}_{k,1}^{(t)}, \dots, \bar{\bm{y}}_{k,L}^{(t)}] \in \mathbb{C}^{M\times L}$ as the contribution from the UE $k$ across the first $L$ sub-blocks of the $t$-th frame, we can express $\bar{\bm{Y}}_{k}^{(t)}$ as:
\begin{equation} \label{eq: pilots in first L block}
    \bar{\bm{Y}}_{k}^{(t)} \triangleq \bm{A}_k^{(t,0)} \bm{V}^{(t)} + \bar{\bm{Z}}_{k}^{(t)} ,
\end{equation}
where $\bm{V}^{(t)} = [\widetilde{\bm{v}}^{(t)}_1, \dots, \widetilde{\bm{v}}^{(t)}_L] \in \mathbb{C}^{(N_r+1)\times L}$ and $\bar{\bm{Z}}_{k}^{(t)} = [\bar{\bm{z}}_{k, 1}^{(t)}, \dots, \bar{\bm{z}}_{k, L}^{(t)}] \in \mathbb{C}^{M\times L}$. 

Note that there are $(M+N_r)K + MN_r$ unknown channel coefficients in
$\bm{G}^{(t,0)}$, $\bm{h}_{r,k}^{(t,0)}$'s, and $\bm{h}_{d,k}^{(t,0)}$'s for $k
= 1,\dots,K$. Since the number of RIS elements $N_r$ is often very large
(possibly hundreds), estimating the channels perfectly would lead to a significant pilot overhead. It is therefore advantageous to bypass explicit
channel estimation and to directly design the beamformers and the RIS reflection
patterns based on the received pilots using a neural network \cite{9914567}.

Furthermore, the time-varying channels across different frames are often highly
correlated, meaning that the historical channel information is valuable for
predicting forthcoming channel states.  It is therefore advantageous to use a
sensing strategy that accounts for the historical channel observations
across the frames, so it can focus on the region of interest to further save 
the pilot overhead, as described next. 

\section{Active Sensing for Beam Tracking}\label{Active Sensing for Tracking}
To exploit the temporal correlations of the time-varying channel, this paper proposes an \textit{active sensing} strategy for designing the RIS sensing vectors (in the first $L$ sub-blocks within the pilot stage of each frame) adaptively across frames based on the pilots received in previous frames. 
Within each frame, we also design 
the downlink data transmission RIS reflection coefficients after $L$ sub-blocks, then with $\tau_w$ additional pilots in the $(L+1)$-th sub-block, design the downlink AP beamformers.  
The detailed process is explained below.

In the pilot stage of the $t$-th transmission frame, the RIS sensing vectors $\{\bm{v}^{(t)}_\ell\}^L_{\ell=1}$ are designed based on the channel observations received prior to the $t$-th frame as:
\begin{equation}
\{\bm{v}^{(t)}_\ell\}^L_{\ell=1} =\mathcal{F}^{(t)}\left(\{\{\bar{\bm{Y}}_{k}^{(j)}\}_{k=1}^{K}\}_{j=1}^{t-1} \right) , \label{eq: UL sensing scheme}
\end{equation}
where the elements of the output RIS sensing vector $\bm{v}^{(t)}_\ell$ should satisfy the unit modulus constraint, i.e., $|[\bm{v}^{(t)}_\ell]_i| = 1$. Here, we refer to $\mathcal{F}^{(t)}: \mathbb{C}^{M\times LK(t-1)}  \rightarrow  \mathbb{C}^{N_r L}$ as the uplink active sensing scheme in the $t$-th frame. Since there are no channel observations before the $1$-st transmission frame, we let $\{\bm{v}^{(1)}_\ell\}^L_{\ell=1} = \mathcal{F}^{(1)}(\varnothing, \varnothing)$, where $\varnothing$ denotes that $\mathcal{F}^{(1)}(\cdot, \cdot)$ takes no input and outputs some random initial $\{\bm{v}^{(1)}_\ell\}^L_{\ell=1}$.

By utilizing the RIS sensing vectors $\{\bm{v}^{(t)}_\ell\}^L_{\ell=1}$, the AP collects a sequence of pilots as in \eqref{eq:pilots_1} in the $t$-th transmission frame. Subsequently, the AP utilizes all the pilots received so far to design the downlink RIS reflection coefficients $\bm{w}^{(t)}$ for the subsequent data transmission within the $t$-th frame:
\begin{equation}
\bm{w}^{(t)} = \mathcal{G}^{(t)}\left( \{\{\bar{\bm{Y}}_{k}^{(j)}\}_{k=1}^{K}\}_{j=1}^{t} \right) , \label{eq: downlink beam alignment scheme}
\end{equation} 
where the element of the output RIS reflection coefficients $\bm{w}^{(t)}$ should again satisfy the unit modulus constraint, i.e., $|[\bm{w}^{(t)}]_i| = 1$. We also refer to $\mathcal{G}^{(t)}: \mathbb{C}^{M\times LKt} \rightarrow \mathbb{C}^{N_r}$ as the downlink beam alignment scheme in the $t$-th frame.

Next, we fix the RIS reflection coefficients as $\bm{w}^{(t)}$ and design the data transmission beamforming matrix $\bm{B}^{(t)}$ at the AP based on the pilot in the $(L+1)$-th sub-block.
To reduce searching space, this paper leverages the following beamforming structure based on uplink-downlink duality \cite{6832894}: 
\begin{equation} \label{beamforming structure}
\bm{b}_{k}^{(t)} \!\! = \!\! \sqrt{p_k^{(t)}} \!\frac{(\bm{I}_M \!+ \! \sum_{i \neq k} \frac{\lambda_i^{(t)}}{\sigma^2_d} \bm{h}_{c,i}^{(t,0)} (\bm{h}_{c,i}^{(t,0)})^{\sf{H}} )^{-1} \bm{h}_{c,k}^{(t,0)}}{\|(\bm{I}_M \!+ \! \sum_{i \neq k} \frac{\lambda_i^{(t)}}{\sigma^2_d} \bm{h}_{c,i}^{(t,0)} (\bm{h}_{c,i}^{(t,0)})^{\sf{H}} )^{-1} \bm{h}_{c,k}^{(t,0)}\|_2} ,
\end{equation} 
where $\bm{h}_{c,k}^{(t,0)} \triangleq \bm{A}_k^{(t,0)} \widetilde{\bm{w}}^{(t)} \in \mathbb{C}^M$ denotes the low-dimensional effective channel between the AP and the UE $k$ in the pilot stage ($0$-th block) of the $t$-th frame. 
In \eqref{beamforming structure}, $p_k^{(t)} \geq 0$ and $\lambda_k^{(t)} \geq 0$ respectively represent for the primal downlink power and the virtual uplink power of the UE $k$ in the $t$-th frame, satisfying the power constraints $\sum_{k=1}^K p_k^{(t)}=\sum_{k=1}^K \lambda_k^{(t)}=P_d$. Accordingly, the beamforming vector design is equivalent to finding the optimal power allocations as follows:
\begin{equation}
    \{\bm{p}^{(t)}, \bm{\lambda}^{(t)}\}  =  \mathcal{H}^{(t)}\left( \{\bar{\bm{y}}_{k,L+1}^{(t)}\}_{k=1}^{K} \right), \label{eq: downlink power allocation scheme}
\end{equation} 
where $\bm{p}^{(t)} \! \triangleq \! [p_1^{(t)},\dots,p_K^{(t)}]^{\sf{T}} \!\in\! \mathbb{R}^K$, $\bm{\lambda}^{(t)} \!\triangleq\! [\lambda_1^{(t)},\dots ,\lambda_K^{(t)}]^{\sf{T}} \!\in \mathbb{R}^K$, and $\mathcal{H}^{(t)}: \mathbb{C}^{MK} \rightarrow \mathbb{R}^{K\times K}$ is the downlink power allocation scheme in the $t$-th frame.

In \eqref{beamforming structure}, we note that the effective channel $\bm{h}_{c,k}^{(t,0)}$ is needed to design the beamforming vectors. Here, since $\bm{h}_{c,k}^{(t,0)}$ is a low-dimensional channel vector, its estimation is straightforward when the RIS reflection coefficients $\widetilde{\bm{w}}^{(t)}$ are fixed.  Specifically, this paper simply estimates $\bm{h}_{c,k}^{(t,0)}$ based on the pilot in the $(L+1)$-th sub-block as follows. 
The decorrelated received pilot $\bar{\bm{y}}_{k, L+1}^{(t)}$ from the UE $k$
in the $(L+1)$-th sub-block is 
$\bar{\bm{y}}_{k, L+1}^{(t)}  = \bm{A}_k^{(t,0)} \widetilde{\bm{w}}^{(t)} + \bar{\bm{z}}_{k, L+1}^{(t)} \triangleq \bm{h}_{c,k}^{(t,0)} + \bar{\bm{z}}_{k, L+1}^{(t)}$.
So approximately, the effective low-dimensional channel $\check{\bm{h}}_{c,k}^{(t,0)} \in \mathbb{C}^{M}$ is simply
\begin{equation} \label{eq: LS}
    \check{\bm{h}}_{c,k}^{(t,0)} = \bar{\bm{y}}_{k, L+1}^{(t)} . 
\end{equation}

We can now formulate the multiuser beam tracking problem for maximizing the minimum downlink user rate as that of designing the uplink active sensing scheme $\mathcal{F}^{(t)}$, the downlink beam alignment scheme $\mathcal{G}^{(t)}$, and the downlink power allocation scheme $\mathcal{H}^{(t)}$ to maximize:
\begin{equation} \label{eq:optimization}
    \begin{array}{ll}
        (\mathrm{P}2)\! : \;\; \underset{\substack{\mathcal{F}^{(t)}(\cdot), \
        \mathcal{G}^{(t)}(\cdot), \\ \mathcal{H}^{(t)}(\cdot)}}{\operatorname{maximize}} & \mathbb{E}\left[\displaystyle \frac{1}{N}\sum_{n=1}^N \left(\min _k R_k^{(t,n)} \right)\right] \\
        \ \ \ \ \ \ \ \ \ \ \mathrm{subject\ to} & \eqref{eq: UL sensing scheme} , \ \eqref{eq: downlink beam alignment scheme} , \ \eqref{beamforming structure}, \ \eqref{eq: downlink power allocation scheme} , \ \eqref{eq: LS} , 
    \end{array}
\end{equation}
where the expectation is taken over the channel and noise distributions, and $N$ is the number of data blocks in every transmission frame. The objective function is chosen to ensure that all UEs meet a downlink rate constraint during all $N$ blocks of data transmissions. 

Analytically solving this variational optimization problem $(\mathrm{P}2)$ is challenging, because it involves a joint optimization over the high-dimensional mappings $\mathcal{F}^{(t)}$, $\mathcal{G}^{(t)}$, and $\mathcal{H}^{(t)}$. Moreover, since the input dimensions of the functions $\mathcal{F}^{(t)}$, $\mathcal{G}^{(t)}$, and $\mathcal{H}^{(t)}$ increase with the number of tracking frames, finding a scalable solution analytically is nearly impossible.

To address the challenge of solving problem $(\mathrm{P}2)$, we propose to utilize deep neural network as a powerful function approximator \cite{liang2017deep} to parameterize the functions $\mathcal{F}^{(t)}$, $\mathcal{G}^{(t)}$, and $\mathcal{H}^{(t)}$. In this way, the computational complexity of the optimization is transferred to the neural network training process \cite{9914567}. 
The key is to identify a neural network architecture capable of summarizing historical channel information across different frames in designing an optimal sensing strategy for beam tracking, as well as in performing effective interference management within each frame.

\section{Learning-Based Active Sensing Framework for Multiuser Beam Tracking}\label{Proposed Data-Driven Approach}
This paper proposes a learning-based active sensing framework that optimizes both the uplink channel sensing and downlink data transmission strategies for effective multiuser beam tracking. Specifically, we employ a GNN to model the spatial relation between the UEs and the RIS, while leveraging the graphic structure to efficiently manage interference among the UEs. 
Furthermore, we utilize a LSTM-based RNN to capture the temporal correlations in the channel dynamics in order to exploit the historical channel information in designing active sensing and beam tracking strategies.

\begin{figure*}[t]
	\includegraphics[scale=0.68]{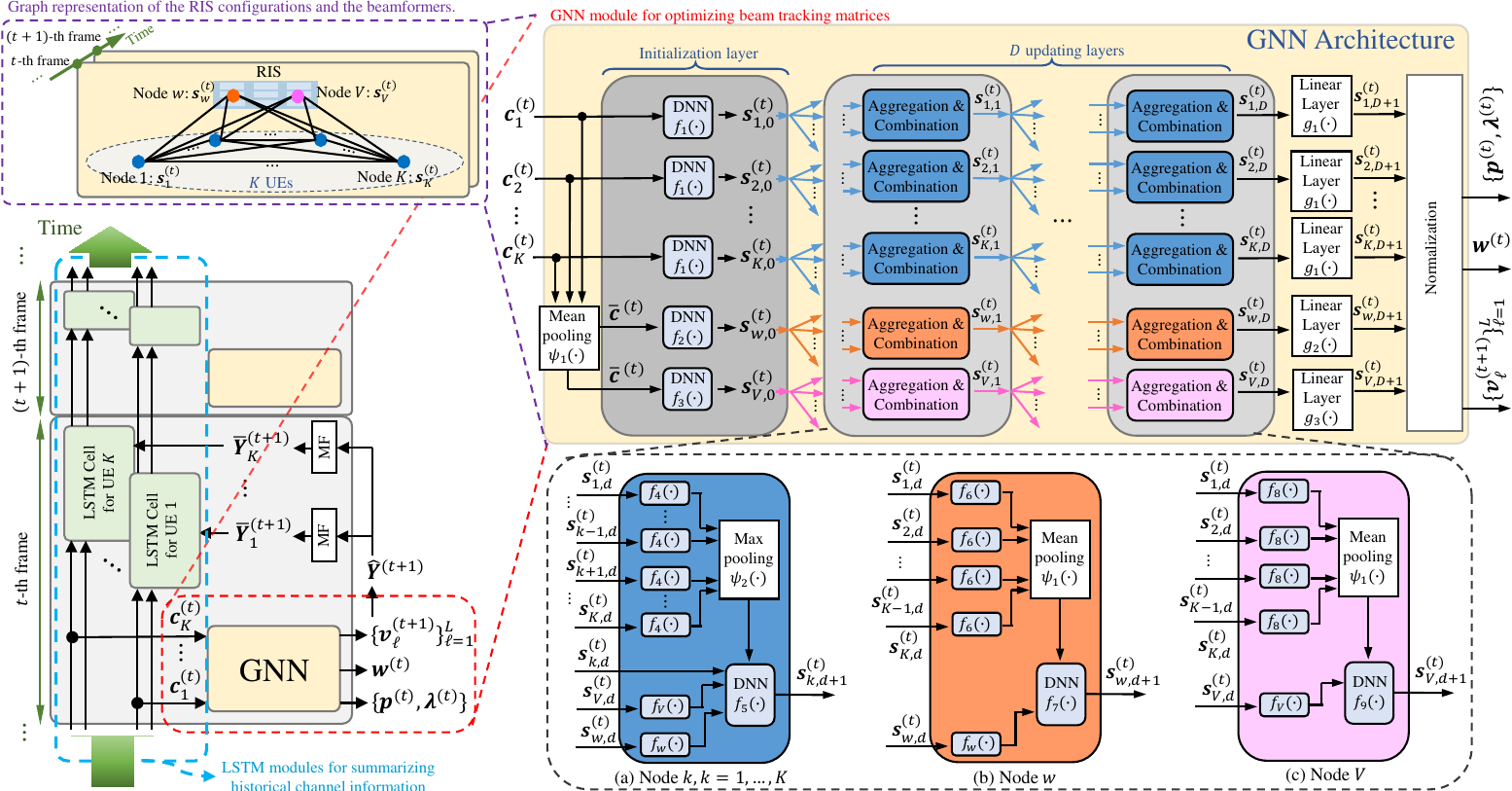}
	\centering
	\caption{Overall graph neural network architecture with an initialization layer, $D$ updating layers, and a final normalization layer.}
	\label{fig:GNN_structure}
\end{figure*}

\subsection{GNN for Interference Management}

In an RIS-assisted multiuser mobile communication systems, a primary challenge 
lies in managing the interference among UEs. Interference management requires
coordination between the downlink beamformers and the RIS reflection
coefficients, which need to be designed based on the received pilots through
appropriate RIS sensing vectors. 

This paper proposes to use a GNN to map the effective CSI information to
the optimized uplink sensing and downlink reflection coefficients at the RIS, as
well as the downlink power allocations at the AP for each transmission frame. A key
motivation for employing a GNN is that it captures the permutation invariant
and equivariant properties of the beam tracking problem. Specifically, the
optimal RIS reflection coefficients should remain unchanged regardless of user
ordering (i.e., permutation invariant). Likewise, any permutation of user index
labels should permute the AP beamformers correspondingly (i.e., permutation
equivariant).  These properties are difficult to learn by a fully connected
neural network, but are embedded in the architecture of a GNN
\cite{ravanbakhsh2017equivariance} and have been successfully applied in the
RIS setting \cite{jiang2021learning}.

The spatial relation among the UEs and the RIS are modeled based on graph representations of the RIS configurations and AP beamformers as shown in Fig.~\ref{fig:GNN_structure}. The graph consists of $K+2$ nodes, with nodes $1$ to $K$ correspond to the power allocations for UE $1$ through UE $K$, and the nodes labeled $V$ and $w$ correspond to the uplink sensing and downlink reflection coefficients at the RIS, respectively. Each node $\xi$, $\xi \in \{w, V, 1,\dots K\}$, is associated with a representation vector denoted by $\bm{s}^{(t)}_\xi$. Specifically, the vector $\bm{s}^{(t)}_w$ is used for designing the downlink RIS reflection coefficients $\bm{w}^{(t)}$, the vector $\bm{s}^{(t)}_V$ is used for designing the RIS sensing vectors $\{\bm{v}^{(t+1)}_\ell\}^L_{\ell=1}$, and the vector $\bm{s}^{(t)}_k$ is used for designing the power allocations $\{p_k^{(t)}, \lambda_k^{(t)}\}$ of the beamformer $\bm{b}^{(t)}_k$. We design the graph with user nodes $1$ to $K$ being fully connected, and establish edges from RIS nodes $V$ and $w$ to all user nodes, respectively. Here, the RIS nodes $V$ and $w$ have different representation vectors, because sensing and data transmission have very different functionalities.


The overall GNN aims to learn the graph representation vectors $\bm{s}^{(t)}_\xi$ through an initialization layer, $D$ updating layers, and a final layer that includes normalization. As shown in Fig.~\ref{fig:GNN_structure}, the GNN first encodes all the useful information about each node in the $t$-th transmission frame into a corresponding representation vector $\bm{s}^{(t)}_{\xi,0}$ according to:
\begin{subequations} \label{eq: GNN 1}
    \begin{align}
        \bm{s}^{(t)}_{k,0} & = f_1 (\bm{c}^{(t)}_k) \ , \ k = 1,\dots,K ,\\
        \bm{s}^{(t)}_{w,0} & = f_2 (\psi_1 (\bm{c}^{(t)}_1,\dots , \bm{c}^{(t)}_K)) ,\\
        \bm{s}^{(t)}_{V,0} & = f_3 (\psi_1 (\bm{c}^{(t)}_1,\dots , \bm{c}^{(t)}_K))  ,
    \end{align}
\end{subequations}
where $\psi_1(\cdot)$ is the element-wise mean pooling function, and $\bm{c}^{(t)}_k$'s are the state information vectors that contain all the useful channel information of the corresponding mobile UEs in the $t$-th frame. Here, $f_1(\cdot)$, $f_2(\cdot)$, and $f_3(\cdot)$ represent fully connected neural networks with the rectified linear unit (ReLU) activation function employed in all the dense layers.

To capture the spatial structure between the RIS and UEs, the initial graph representation vectors $\bm{s}^{(t)}_{\xi,0}$ undergo $D$ iterations of updates, each utilizing the output of the preceding layer as input. More specifically, the updates for the RIS node $w$ and $V$ incorporate their current representations as well as the representations of all UE nodes, while the representation vector of the UE node $k$ is updated based on its own prior representation and the representations of all other nodes in the network. These update procedures maintain the GNN's permutation invariance and equivariance, and therefore allows more efficient training. The updating rules of the $d$-th updating layer are given as:
\begin{subequations} \label{eq: GNN 2}
    \begin{align}
        \bm{s}^{(t)}_{k,d+1} & = f_5 (\bm{s}^{(t)}_{k,d}, f_V(\bm{s}^{(t)}_{V,d}), f_w(\bm{s}^{(t)}_{w,d}), \psi_2(\{f_4(\bm{s}^{(t)}_{\xi,d})\}_{\forall \xi \neq k})) ,\\
        \bm{s}^{(t)}_{w,d+1} & = f_7 (f_w(\bm{s}^{(t)}_{w,d}), \psi_1(f_6(\bm{s}^{(t)}_{1,d}),\dots , f_6(\bm{s}^{(t)}_{K,d})) ,\\
        \bm{s}^{(t)}_{V,d+1} & = f_9 (f_V(\bm{s}^{(t)}_{V,d}), \psi_1(f_8(\bm{s}^{(t)}_{1,d}),\dots , f_8(\bm{s}^{(t)}_{K,d})) .
    \end{align}
\end{subequations}
Here, $\psi_2(\cdot)$ is the element-wise max pooling function, while $f_4(\cdot),f_5(\cdot),f_6(\cdot),f_7(\cdot),f_8(\cdot),f_9(\cdot), f_w(\cdot), f_V(\cdot)$ are fully connected neural networks with ReLU activation function applied in each of their dense layers.

After $D$ layers of iteration, the updated representation vectors $\bm{s}^{(t)}_{\xi,D}$ would contain sufficient information to design the beam tracking matrices. To obtain the correct output dimensions, we pass $\bm{s}^{(t)}_{k,D}$'s, $\bm{s}^{(t)}_{w,D}$, and $\bm{s}^{(t)}_{V,D}$ through linear layers $g_1(\cdot),g_2(\cdot),g_3(\cdot)$ with $2$, $2N_r$, and $2N_rL$ fully connected units, respectively. Specifically, $\bm{s}^{(t)}_{k,D+1} = g_1 (\bm{s}^{(t)}_{k,D}) \in \mathbb{R}^{2}$, $\bm{s}^{(t)}_{w,D+1} = g_2 (\bm{s}^{(t)}_{w,D}) \in \mathbb{R}^{2N_r}$, and $\bm{s}^{(t)}_{V,D+1} = g_3 (\bm{s}^{(t)}_{V,D}) \in \mathbb{R}^{2N_rL}$. The final representation vectors $\bm{s}^{(t)}_{\xi,D+1}$ are then normalized to produce the RIS sensing vectors $\{\bm{v}^{(t+1)}_\ell\}^L_{\ell=1}$, downlink RIS reflection coefficients $\bm{w}^{(t)}$, and power allocations $\{\bm{p}^{(t)}, \bm{\lambda}^{(t)}\}$, respectively, while ensuring that the unit modulus constraints on $\bm{w}^{(t)}$, $\bm{v}_{\ell}^{(t+1)}$'s and the power constraints on $\{\bm{p}^{(t)}, \bm{\lambda}^{(t)}\}$ are satisfied. 

\begin{figure}[t]
	\includegraphics[scale=0.48]{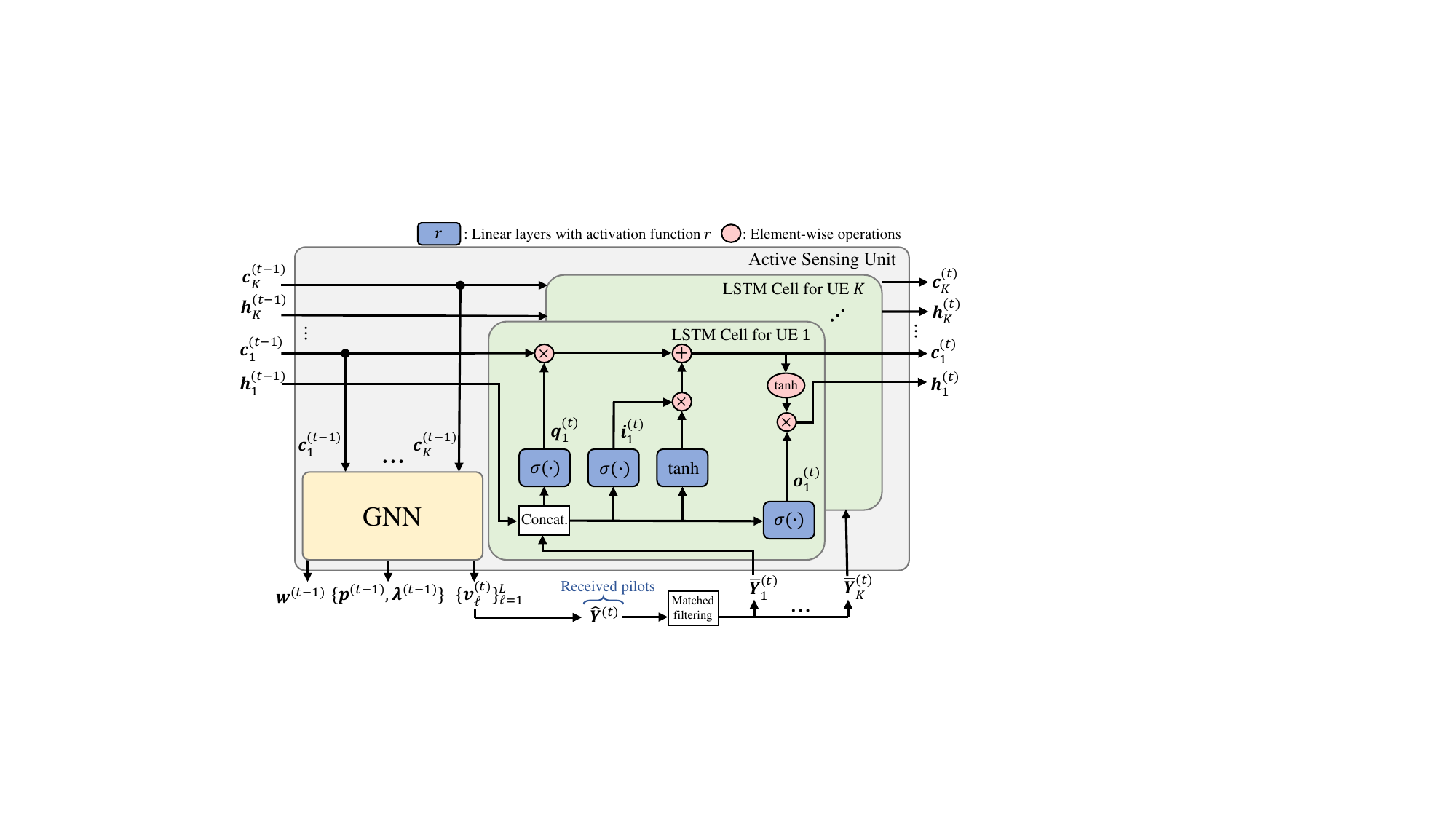}%
	\centering
	\caption{Active sensing unit for multiuser beam tracking with RIS.} 
	\label{fig:ActiveSensingUnit}
\end{figure}

\subsection{LSTM for Exploiting Temporal Channel Correlation}

This paper proposes to use an LSTM-based RNN to capture the temporal correlations of the time-varying channel by automatically summarizing the channel information into state information vectors $\bm{c}^{(t)}_k$ in \eqref{eq: GNN 1}, which are further processed by the GNN to design the beam tracking strategies. As shown in Fig.~\ref{fig:ActiveSensingUnit}, $K$ copies of the LTSM cells with shared weights are deployed to summarize the state information for the $K$ mobile UEs, respectively. In the $t$-th frame, the $k$-th LSTM cell takes new pilots $\bar{\bm{Y}}_{k}^{(t)}$ from UE $k$ to update its hidden state vector $\bm{h}_k^{(t)}$ and cell state vector $\bm{c}_k^{(t)}$ according to \cite{6795963}:
\begin{subequations} \label{eq: LSTM 1}
    \begin{align}
        \bm{c}^{(t)}_k & =\bm{q}_k^{(t)} \circ \bm{c}_k^{(t-1)}+\bm{i}_k^{(t)} \circ \tanh (\bm{u}_c(\bar{\bm{Y}}_{k}^{(t)})+\bm{w}_c(\bm{h}_k^{(t-1)})) , \\
        \bm{h}_k^{(t)} & = \bm{o}_k^{(t)} \circ \tanh (\bm{c}_k^{(t)}) ,
    \end{align}
\end{subequations}
where $\bm{q}_k^{(t)}$, $\bm{i}_k^{(t)}$, and $\bm{o}_k^{(t)}$ are the activation vectors of the forget gate, input gate and output gate within the $k$-th LSTM cell, respectively. The updating rules for different gates are given as $\bm{r}_k^{(t)} =\sigma (\bm{u}_r(\bar{\bm{Y}}_{k}^{(t)})+\bm{w}_r(\bm{h}_k^{(t-1)}))$, where $ r \in \{q,i,o\}$, and $\sigma(\cdot)$ is the element-wise sigmoid function. Here, $\bm{u}_c$, $\bm{u}_q$, $\bm{u}_i$, $\bm{u}_o$, $\bm{w}_c$, $\bm{w}_q$, $\bm{w}_i$, and $\bm{w}_o$ are fully connected layers, with the weights of each being shared across $K$ LSTM cells.

\begin{figure*}[t]
	\includegraphics[scale=0.76]{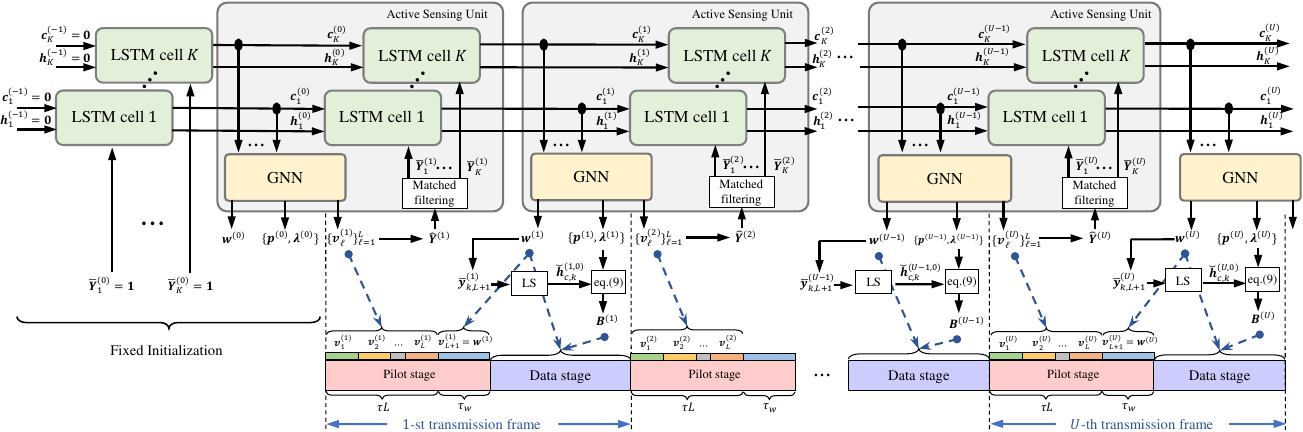}%
	\centering
	\caption{Proposed active sensing framework for multiuser beam tracking with RIS. During the training phase, we concatenate $U$ active sensing units corresponding to $U$ transmission frames. Once trained and deployed, we reuse the active sensing units for a potentially infinite number of transmission frames.}
	\label{fig:LSTMGNN_Training}
\end{figure*}

\subsection{Active Sensing Framework for Beam Tracking}

The overall active sensing framework is shown in
Fig.~\ref{fig:LSTMGNN_Training}. It works as follows. 

\subsubsection{RIS Sensing Vector Design}
In the pilot stage of the $(t-1)$-th frame, the LSTM cell states $\bm{c}_k^{(t-1)}$ that encapsulate the temporal channel correlations for the $K$ mobile UEs based on the previous frames are fed into a GNN. The GNN then outputs the RIS sensing vectors $\{\bm{v}^{(t)}_\ell\}^L_{\ell=1}$ for the next round of CSI acquisition (i.e., pilot stage of the $t$-th frame). We observe empirically that it is also possible to replace the LSTM cell state $\bm{c}_k^{(t-1)}$ with the hidden state $\bm{h}_k^{(t-1)}$ as the GNN input. 

\subsubsection{State Vector Update}
In the $t$-th frame, $K$ LSTM cells with shared weights take in the most recent channel observations $\bar{\bm{Y}}_{k}^{(t)}$, alongside the preceding cell states $\bm{c}_k^{(t-1)}$ and hidden states $\bm{h}_k^{(t-1)}$, to generate the new cell states $\bm{c}_k^{(t)}$ and hidden states $\bm{h}_k^{(t)}$. By convention, we initialize $\bm{c}_k^{(-1)}=\bm{h}_k^{(-1)}=\bm{0}$, and $\bar{\bm{Y}}_{k}^{(0)}=\bm{1}$ for all $k$ in $\{0,\dots,K\}$.

\subsubsection{Downlink Beam Alignment}
Utilizing the updated LSTM cell states $\bm{c}_k^{(t)}$ as input, the proposed GNN computes the downlink RIS reflection coefficients $\bm{w}^{(t)}$ and power allocations $\{\bm{p}^{(t)}, \bm{\lambda}^{(t)}\}$. The AP then estimates the effective low-dimensional channels $\bm{h}_{c,k}^{(t,0)}$ based on the additional $\tau_w$ pilot symbols transmitted via the reflection of the $(L+1)$-th RIS sensing vector $\bm{v}_{L+1}^{(t)} = \bm{w}^{(t)}$. With the estimated low-dimensional channel $\check{\bm{h}}_{c,k}^{(t,0)}$ and designed power allocations $\{\bm{p}^{(t)}, \bm{\lambda}^{(t)}\}$, the downlink beamforming vectors $\bm{b}^{(t)}_k$ for subsequent data transmissions are computed via \eqref{beamforming structure}.

\subsection{Neural Network Training and Inference}

The active sensing units are concatenated 
to form a deep neural network as shown in
Fig.~\ref{fig:LSTMGNN_Training}, with each unit corresponding to a transmission
frame. The neural network weights are tied together across the stages. 
By training the concatenated neural network, the active sensing unit is
able to explore the temporal channel correlation from the input pilot sequence.
The proposed architecture is trained offline in an unsupervised fashion.
Specifically, we employ Adam optimizer \cite{kingma2014adam} to minimize the
negative of the average minimum downlink rate over $U$ frames as
$-\frac{1}{U}\sum_{t=1}^{U}\mathbb{E}[\frac{1}{N}\sum_{n=1}^N (\min _k
R_k^{(t,n)} )]$, where the expectation is approximated by the empirical average
over the training set. In this way, the uplink active sensing scheme
$\mathcal{F}^{(t)}$, the downlink beam alignment scheme $\mathcal{G}^{(t)}$,
and the downlink power allocation scheme $\mathcal{H}^{(t)}$ are jointly
designed to optimize the beam tracking performance. In practice, we choose the
number of the concatenated active sensing units $U$ to be sufficiently large so
that the temporal correlations of the time-varying channel can be learned
through the periodically received pilots. Once the neural network is trained,
the trained active sensing unit can be reused for a potentially infinite number
of frames, since the LSTM cell learns to keep updating the channel
information based on the latest received pilots.

In the simulations, we observe that it is possible to further improve the performance of directly using the power allocations $\{\bm{p}^{(t)}, \bm{\lambda}^{(t)}\}$ designed by neural network. 
This is primarily because the proposed active sensing framework only has moderate size of the hidden neurons, which restricts its expressive capacity. 
It is possible to adopt a refined approach using the fact that the effective low-dimensional channels $\bm{h}_{c,k}^{(t,0)}$ can be easily estimated according to \eqref{eq: LS}. By employing a fixed-point iterations method \cite{9066923,5762643} which takes the estimated $\check{\bm{h}}_{c,k}^{(t,0)}$'s as input, we can analytically determine the optimal power allocations. 
This method allows for a more precise adjustment of power allocations, ensuring that the beamforming vectors computed therefrom are better optimized.

More specifically, the optimal power allocation $\bm{\lambda}^{(t)}_*$ is the unique positive solution of the following set of fixed-point equations, for $k=1,\cdots, K$, 
\begin{equation}
    \begin{aligned}
        & [\bm{\lambda}^{(t)}_*]_k = \frac{\tau^{(t)}_*}{f(\bm{\lambda}^{(t)}_*, \check{\bm{h}}_{c,k}^{(t,0)}, \{\check{\bm{h}}_{c,i}^{(t,0)}\}_{\forall i \neq k})} \\
        & \triangleq \frac{\tau^{(t)}_*}{(\check{\bm{h}}_{c,k}^{(t,0)})^{\sf{H}}(\sum_{i \neq k} [\bm{\lambda}^{(t)}_*]_i \check{\bm{h}}_{c,i}^{(t,0)} (\check{\bm{h}}_{c,i}^{(t,0)})^{\sf{H}}+\sigma_d^2 \bm{I}_M)^{-1} \check{\bm{h}}_{c,k}^{(t,0)}} , \label{eq: fixed point euqation 1}
    \end{aligned}
\end{equation}
where $\tau^{(t)}_* \! = \! P_d / \sum_{k=1}^K (f(\bm{\lambda}^{(t)}_*, \check{\bm{h}}_{c,k}^{(t,0)}, \{\check{\bm{h}}_{c,i}^{(t,0)}\}_{\forall i \neq k}))^{-1}$. 

Further, the optimal power allocation $\bm{p}^{(t)}_*$ is obtained as:
\begin{equation} \label{eq: fixed point euqation 3}
\bm{p}^{(t)}_*=\left(\bm{I}_K-\tau^{(t)}_* \bm{D}^{(t)} \bm{F}^{(t)}\right)^{-1} \tau^{(t)}_* \sigma_d^2 \bm{D}^{(t)} \bm{1}_K ,
\end{equation}
where $\bm{D}^{(t)} = \text{diag}\left(\frac{1}{|(\check{\bm{h}}_{c,1}^{(t,0)})^{\sf{H}} \widetilde{\bm{b}}^{(t)}_{1}|^2},\dots,\frac{1}{|(\check{\bm{h}}_{c,K}^{(t,0)})^{\sf{H}} \widetilde{\bm{b}}^{(t)}_{K}|^2}\right)$ and $[\bm{F}^{(t)}]_{k,i} = \frac{1}{\left|\left(\check{\bm{h}}_{c,k}^{(t,0)}\right)^{\sf{H}} \widetilde{\bm{b}}^{(t)}_{i}\right|^2}$, if $k \neq i$ and $0$ otherwise. Here, $\widetilde{\bm{b}}^{(t)}_{k}$'s denote the beamforming directions which can be computed based on $\bm{\lambda}^{(t)}_*$ and $\check{\bm{h}}_{c,k}^{(t,0)}$'s via \eqref{beamforming structure}.

Note however, in order to preserve the differentiability of gradients during back-propagation, we cannot employ the fixed-point iterations for computing power allocations in the neural network's training phase. Only in the inference phase of the proposed active sensing framework, we replace the power allocations $\{\bm{p}^{(t)}, \bm{\lambda}^{(t)}\}$ designed by the GNN with $\{\bm{p}^{(t)}_*, \bm{\lambda}^{(t)}_*\}$ analytically determined through fixed-point iterations to obtain the best performance.

\section{Numerical Results and Interpretations} \label{Numerical Results and Interpretations}
\subsection{Simulation Setup}
We consider an RIS-assisted MU-MISO communication system as illustrated in Fig.~\ref{fig: Simulation layout}, consisting of an AP with $8$ antennas and an RIS with $100$ passive elements. In the Cartesian coordinate system shown in Fig.~\ref{fig: Simulation layout}, the $(x,y,z)$-coordinates of the AP and the RIS are $(100\text{m}, -100\text{m}, 0\text{m})$ and $(0\text{m}, 0\text{m}, 0\text{m})$, respectively. We assume that the RIS is equipped with a $10\times10$ uniform rectangular array placed on the $(y,z)$-plane, and the AP has a uniform linear array parallel to the $x$-axis. The system is assumed to operate at a carrier frequency $f_c = 1\text{GHz}$ with $10$MHz bandwidth. The uplink pilot transmit power $P_u$ and the downlink data transmit power $P_d$ are set to be $5$dBm and $10$dBm, and the noise spectrum density at the AP and UE are set to be $N_{0,u} = -154$dBm/Hz and $N_{0,d} = -160$dBm/Hz, respectively.

\subsubsection{UE Mobility Model}
\begin{figure*}[t]
\minipage{0.35\textwidth}
  \includegraphics[width=\linewidth]{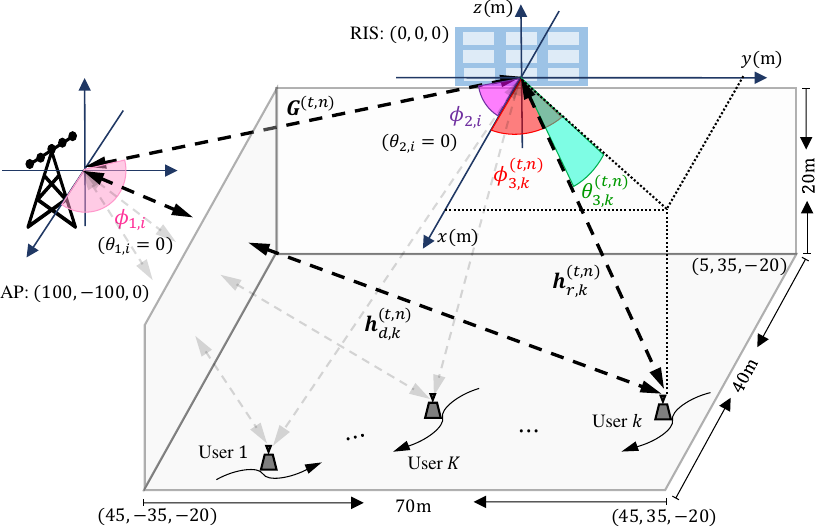}
  \caption{System geometries for simulations.}\label{fig: Simulation layout} 
\endminipage\hfill
\minipage{0.37\textwidth}
  \includegraphics[width=\linewidth]{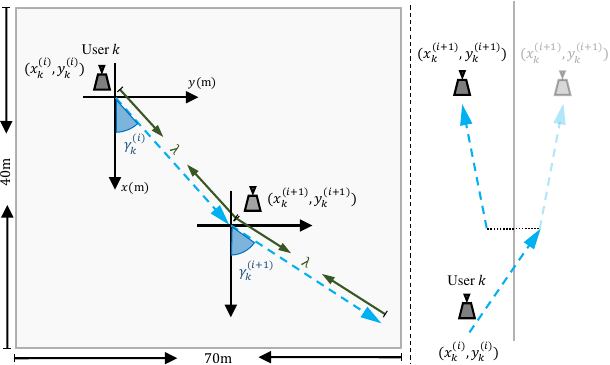}
  \caption{Illustration of the UE's mobility model.}\label{fig: mobility model}
\endminipage\hfill
\minipage{0.22\textwidth}%
  \includegraphics[width=\linewidth]{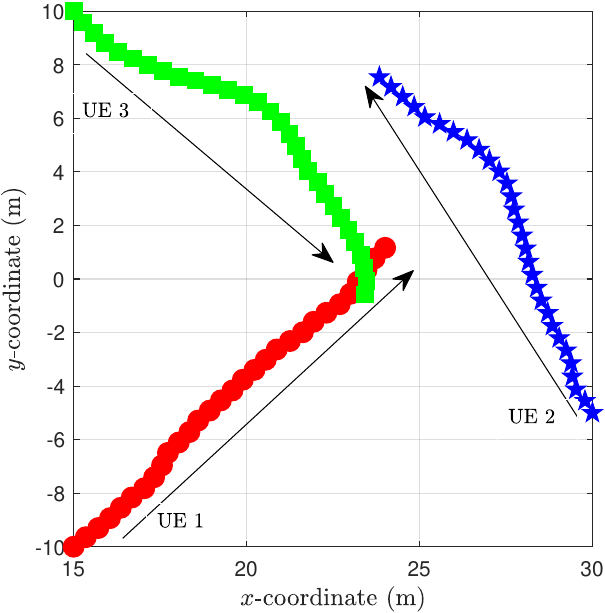}
  \caption{Trajectories of UEs.}\label{fig: Trajectory} 
\endminipage
\end{figure*}

As shown in Fig.~\ref{fig: Simulation layout}, there are $K=3$ mobile UEs located within a rectangular service area $[5\text{m},45\text{m}]\times[-35\text{m},35\text{m}]$ on the $(x,y)$-plane, with their $z$-coordinates fixed as $-20\text{m}$. Furthermore, as illustrated in Fig.~\ref{fig: mobility model}, the $(x,y)$-coordinates of the mobile UE $k$ in the $(i+1)$-th time block are determined as:
\begin{equation}
    [x_k^{(i+1)}, y_k^{(i+1)}]^{\sf{T}} = [x_k^{(i)},y_k^{(i)}]^{\sf{T}} + [\lambda \cos \gamma_k^{(i)},\lambda \sin \gamma_k^{(i)}]^{\sf{T}} \! ,
\end{equation}
where $\gamma_k^{(i)}$ denotes the moving direction of the UE $k$ in the $i$-th time block and $\lambda$ denotes the UE displacement per block. Here, the initial position of the UEs is uniformly distributed within the service area, i.e., $x_k^{(1)} \sim \mathcal{U}(5\text{m},45\text{m})$ and $y_k^{(1)} \sim \mathcal{U}(-35\text{m},35\text{m})$. To model practical trajectories of the mobile UEs, we assume that $\gamma_k^{(i)}$ is updated at each block as $\gamma_k^{(i+1)} = \gamma_k^{(i)} + \triangle\gamma$, where the initial moving direction $\gamma_k^{(1)} \sim \mathcal{U}(0,2\pi)$ and $\triangle\gamma \sim \mathcal{U}(-10^{\circ}, 10^{\circ})$. We further assume that the mobile UEs move at a constant speed of $v_{\text{UE}} = 10\text{m/s}$, which results in a maximum Doppler frequency $f_{max} = v_{\text{UE}} f_c/c = 33.3\text{Hz}$ with $c=3\times10^8\text{m/s}$ as the speed of light. The UE displacement per block is then calculated as $\lambda = v_{\text{UE}} T_b = v_{\text{UE}}/2 f_{max} = 0.15\text{m}$, where $T_b = 1/2 f_{max}$ is the duration of each block. The mobile UEs are restricted to a $40\text{m}\times70\text{m}$ service area. If a UE crosses this boundary, its coordinates and movement direction are mirrored back into the area, as shown in Fig.~\ref{fig: mobility model}.

\subsubsection{Channel Model}
In the $n$-th block of the $t$-th transmission frame, we assume that the channels from the AP to the mobile UEs are modeled by the Rayleigh fading as:
\begin{equation}
    \bm{h}_{d,k}^{(t,n)}=\beta_{d,k}^{(t,n)}\widetilde{\bm{h}}_{d,k}^{(t,n)} ,
\end{equation}
where $\beta_{d,k}^{(t,n)}$ denotes the path-loss of the direct link between the AP and the UE $k$ modeled in (dB) as $32 + 43.3\log(l_{d,k}^{(t,n)})$, where $l_{d,k}^{(t,n)}$ is the distance of the direct link from the AP to the UE $k$. The variation of the entries of the non-line-of-sight (NLOS) components of the direct link $[\widetilde{\bm{h}}_{d,k}^{(t,n)}]_i$ is modeled by the stationary Gauss-Markov process as follows \cite{va2016beam}:
\begin{equation}
    \left\{
        \begin{aligned}
            [\widetilde{\bm{h}}_{d,k}^{(t,n)}]_{i} &= \rho [\widetilde{\bm{h}}_{d,k}^{(t-1,N)}]_{i} + \sqrt{1 - \rho^2} \mu^{(t-1,N)} ,\ n = 1 , \\ 
            [\widetilde{\bm{h}}_{d,k}^{(t,n)}]_{i} &= \rho [\widetilde{\bm{h}}_{d,k}^{(t,n-1)} ]_{i}+ \sqrt{1 - \rho^2} \mu^{(t,n-1)} , 2\leq n \leq N , 
        \end{aligned}
    \right.
    \label{eq:gauss-markov}
\end{equation}
where the correlation coefficient $\rho = 0.9995$, perturbation term $\mu^{(k,n)} \sim \mathcal{CN}(0, 1)$, and $[\widetilde{\bm{h}}_{d,k}^{(1,1)}]_{i} \sim \mathcal{CN}(0, 1)$. We assume that the RIS is deployed at a location where multiple paths exist between the AP and the RIS, and a LOS channel exists between the RIS and each mobile UE. More specifically, the AP-RIS link $\bm{G}^{(t,n)}$ and the RIS-UE links $\bm{h}_{r,k}^{(t,n)}$ follow the Rician fading model as:
\begin{equation}
    \bm{G}^{(t,n)}\!=\!\beta_{G}^{(t,n)}\! \left(\sqrt{\varepsilon/(1+\varepsilon)}  \sum_i^{N_G} \widehat{\bm{G}}_i +\sqrt{1/(1+\varepsilon)} \ \widetilde{\bm{G}}^{(t,n)}\right),
\end{equation}
\begin{equation}
    \bm{h}_{r,k}^{(t,n)}\!=\!\beta_{r,k}^{(t,n)}\! \left(\sqrt{\varepsilon/(1+\varepsilon)} \ \widehat{\bm{h}}_{r,k}^{(t,n)}+\sqrt{1/(1+\varepsilon)} \ \widetilde{\bm{h}}_{r,k}^{(t,n)}\right) ,
\end{equation}
where $N_G$ is the number of paths between the AP and RIS, and $\varepsilon$ is the Rician factor which is set to be $10$. The path loss (in dB) of the AP-RIS link $\beta_{G}$ and the RIS-UE link $\beta_{r}^{(t,n)}$ are modeled as $28 + 16.9\log(l_{G})$ and $28 + 16.9\log(l_{r,k}^{(t,n)})$, respectively, where $l_{G}$ and $l_{r,k}^{(t,n)}$ are the distance of the corresponding links. Similar to \eqref{eq:gauss-markov}, the distribution of the entries of the NLOS components of the AP-RIS link $[\widetilde{\bm{G}}^{(t,n)}]_{ij}$ and the RIS-UE links $[\widetilde{\bm{h}}_{r,k}^{(t,n)}]_{i}$ are modeled by the same stationary Gauss-Markov process, with $[\widetilde{\bm{G}}^{(1,1)}]_{ij} \sim \mathcal{CN}(0, 1)$ and $[\widetilde{\bm{h}}_{r,k}^{(1,1)}]_{i} \sim \mathcal{CN}(0, 1)$.

The LOS part of the RIS-UE link $\widehat{\bm{h}}_{r,k}^{(t,n)}$ is a function of the UE/RIS locations. Specifically, let $\theta_{3,k}^{(t,n)}$, $\phi_{3,k}^{(t,n)}$ denote the elevation and azimuth angle of arrivals (AoAs) from the UE $k$ to the RIS at the $n$-th block of the $t$-th frame, as shown in Fig.~\ref{fig: Simulation layout}. We can write $\widehat{\bm{h}}_{r,k}^{(t,n)} = \bm{a}_{\text{RIS}}(\theta_{3,k}^{(t,n)}, \phi_{3,k}^{(t,n)})$, where the $u$-th element of the steering vector at RIS is given as \cite{jiang2021learning}:
\begin{equation} \label{RIS steering vector}
    \begin{aligned}
        & {[\bm{a}_{\text{RIS}}(\theta_{3, k}^{(t,n)}, \phi_{3, k}^{(t,n)})]_u} \\
        & =e^{j \frac{2 \pi d_{\text{RIS}}}{\lambda_c}\left(i_1(u,v) \sin (\phi_{3, k}^{(t,n)}) \cos (\theta_{3, k}^{(t,n)})+i_2(u,v) \sin (\theta_{3, k}^{(t,n)})\right)} .
    \end{aligned}
\end{equation}
Here, $\lambda_c$ is the carrier wavelength, $d_{\text{RIS}}$ is the distance between two adjacent RIS elements, $i_1(u, v)=\bmod (u-1, v)$, and $i_2(u, v)=\lfloor\frac{u-1}{v}\rfloor$, where $v$ is the number of columns of the RIS. Without loss of generality, we set $d_{\text{RIS}} = \lambda_c / 2$ and $v=10$ in simulations. As shown in Fig.~\ref{fig: Simulation layout}, the RIS is located at $(0\text{m}, 0\text{m}, 0\text{m})$. Let $(x_k^{(t,n)}, y_k^{(t,n)}, z_k)$ denote the coordinates of the UE $k$ with $z_k=-20\text{m}$, we have $\sin (\phi_{3, k}^{(t,n)}) \cos (\theta_{3, k}^{(t,n)}) = y_k^{(t,n)}/l_{r,k}^{(t,n)}$ and $\sin (\theta_{3, k}^{(t,n)}) = z_k/l_{r,k}^{(t,n)}$.
Similarly, let $\phi_{2,i}$ and $\phi_{1,i}$ denote the $i$-th azimuth angle of departure (AoD) and azimuth AoA from the RIS to the AP, corresponding to the $i$-th path $\widehat{\bm{G}}_i$ between the AP and the RIS. Then we can write $\widehat{\bm{G}}_i = \bm{a}_{\text{AP}}(\phi_{1,i}) \bm{a}_{\text{RIS}}(\theta_{2,i}, \phi_{2,i})^{\sf{H}}$, where $\theta_{2,i} = 0$ as shown in Fig.~\ref{fig: Simulation layout} and the steering vector of the AP is given by:
\begin{equation}
    \bm{a}_{\text{AP}}(\phi_{1,i}) = [1, \cdots, e^{j \frac{2 \pi(M-1) d_{\text{AP}}}{\lambda_c} \cos \left(\phi_{1,i}\right)}]^{\sf{T}} .
\end{equation}
Here, $d_{\text{AP}}$ denotes the distance between two adjacent antenna elements. In simulations, we set $d_{\text{AP}} = \lambda_c / 2$ and generate the AoAs/AoDs according to uniform distributions, i.e., $\phi_{1,i} \sim \mathcal{U}(0,2\pi)$ and $\phi_{1,i} \sim \mathcal{U}(-\pi/2,0)$. 

\subsection{Implementation Details}
The proposed active sensing framework is implemented on Tensorflow \cite{abadi2016tensorflow}. For the LSTM cells, the dimensions of hidden states $\bm{h}_k^{(t)}$, cell states $\bm{c}_k^{(t)}$, and hidden layers $\bm{u}_c$, $\bm{u}_q$, $\bm{u}_i$, $\bm{u}_o$, $\bm{w}_c$, $\bm{w}_q$, $\bm{w}_i$, $\bm{w}_o$ are set to be $512$. We adopt a GNN with $D=2$ updating layers. Within the GNN architecture, the DNNs $f_1(\cdot)$, $f_2(\cdot)$, and $f_3(\cdot)$ are of size $512 \times 1024 \times 512$, and the size of DNNs $f_4(\cdot)$, $f_5(\cdot)$, $f_6(\cdot)$, $f_7(\cdot)$, $f_8(\cdot)$, $f_9(\cdot)$, $f_w(\cdot)$, $f_V(\cdot)$ are $512 \times 512 \times 512$. We concatenate $U=20$ active sensing units to train the proposed framework. In each training epoch, the neural network samples $6400$ training data including random channel and noise realizations. We terminate the training process if the validation loss does not decrease over $30$ consecutive training epochs. An instance of trajectories for $K=3$ mobile UEs is shown in Fig.~\ref{fig: Trajectory}. 

We assume there are $N_G = 8$ paths between the AP and RIS. Each transmission frame contains $N = 3$ blocks. In the pilot stage (the $0$-th block) of each frame, there are $4$ sub-blocks, i.e., $L=3$. The effective downlink channels $\bm{h}_{c,k}^{(t,0)}$ are estimated in the $4$-th sub-block with $\tau_w = 10$ pilots.

\subsection{Benchmark Methods}
\subsubsection*{BCD with Perfect CSI \cite{9087848}}
Given the perfect CSI of the channels $\bm{G}^{(t,1)}$, $\bm{h}_{\mathrm{d},k}^{(t,1)}$, and $\bm{h}_{\mathrm{r},k}^{(t,1)}$, we can find a performance benchmark for the proposed active sensing framework by solving the following optimization problem:
\begin{equation}\label{eq: optimization - p3}
    (\bm{B}^{(t)}_*, \bm{w}^{(t)}_*) = \mathrm{argmax}_{\bm{B}^{(t)}, \bm{w}^{(t)}} \min_k \{R_k^{(t,1)}\} , \tag{P3}
\end{equation}
subject to $\sum_k\|\bm{b}^{(t)}_k\|_2^2 \leq P_d$ and $|[\bm{w}^{(t)}]_n| = 1$ using the block coordinate descent (BCD) algorithm, which iteratively optimizes the beamforming matrix $\bm{B}^{(t)}$ and the downlink RIS reflection coefficients $\bm{w}^{(t)}$. Specifically, in each iteration, we optimize $\bm{B}^{(t)}$ using the fixed-point iterations \eqref{eq: fixed point euqation 1} - \eqref{eq: fixed point euqation 3} with $\bm{w}^{(t)}$ fixed \cite{9066923}, and then optimize $\bm{w}^{(t)}$ using the Riemannian conjugate gradient (RCG) method with fixed $\bm{B}^{(t)}$ \cite{yu2019miso, 9681803}.

\subsubsection*{GNN with Fixed Sensing Scheme \cite{jiang2021learning}}
In the first $L$ sub-blocks of each transmission frame, the AP receives pilots through a fixed set of the sensing vectors $\{\bm{v}_\ell\}^L_{\ell=1}$ which are generated either randomly or learned from channel statistics. The GNN as described in the proposed approach is employed to map the received pilots $\bar{\bm{Y}}_{k}^{(t)}$ to the downlink RIS reflection coefficients $\bm{w}^{(t)}$. Subsequently, $\tau_w$ additional pilot symbols are transmitted to estimate the effective channels $\bm{h}_{c,k}^{(t,0)}$ for power allocation refinement. The sensing vector $\bm{v}_{L+1}^{(t)}$ is set to be $\bm{w}^{(t)}$ in this refinement stage.

\subsubsection*{GNN \& LSTM Without Active Sensing \cite{jiang2021learning,sohrabi2022active,liu2022scalable}}\label{LSTM-Based Design with Fixed Sensing Vectors}
This baseline is similar to the proposed approach, except that the RIS sensing vectors $\{\bm{v}^{(t)}_\ell\}^L_{\ell=1}$ are not actively designed across different transmission frames. Without active sensing, $\bm{v}^{(t)}_\ell$'s can be designed based on the following three schemes: (i) the phases of each $\bm{v}^{(t)}_\ell$ are randomly drawn from a uniform distribution over $[0, 2\pi)$  \cite{jiang2021learning}; (ii) $\bm{v}^{(t)}_\ell$'s are learned from statistics according to channel realizations in the training phase \cite{sohrabi2022active}; (iii) using the current RIS reflection coefficients $\bm{w}^{(t)}$ for the next round of pilot acquisition \cite{liu2022scalable}, i.e., $\bm{v}_\ell^{(t+1)} = \bm{w}^{(t)}$, $\ell=1,\dots,L$. In (i), (ii), and (iii), $\tau_w$ additional pilot symbols are transmitted in the $(L+1)$-th sub-block with $\bm{v}_{L+1}^{(t)} = \bm{w}^{(t)}$ to estimate the effective channels $\bm{h}_{c,k}^{(t,0)}$ for power allocation refinement. 

\subsubsection*{GNN \hspace{-0.1mm}\&\hspace{-0.1mm} LSTM with GNN Directly Designing $\bm{B}^{(t)}$}
This benchmark is similar to the proposed active sensing framework, except that the beamforming structure \eqref{beamforming structure} is not leveraged to design the AP beamforming matrix $\bm{B}^{(t)}$ in the training stage. In this case, instead of outputting the power allocations $\{\bm{p}^{(t)}, \bm{\lambda}^{(t)}\}$, the GNN directly outputs the exact downlink beamforming vectors $\bm{b}^{(t)}_k$ for each user. During inference, $\tau_w$ additional pilot symbols are transmitted in the $(L+1)$-th sub-block with $\bm{v}_{L+1}^{(t)} = \bm{w}^{(t)}$ for power allocation refinement.

\subsubsection*{GNN \& LSTM Without Power Allocation Refinement}
This benchmark is similar to the proposed active sensing framework but without the power allocation refinement stage. That is, it computes the downlink beamforming matrix $\bm{B}^{(t)}$ using the power allocations $\{\bm{p}^{(t)}, \bm{\lambda}^{(t)}\}$ produced by the GNN, without replacing it with $\{\bm{p}_*^{(t)}, \bm{\lambda}_*^{(t)}\}$ as computed in \eqref{eq: fixed point euqation 1}-\eqref{eq: fixed point euqation 3}. In this benchmark, the RIS sensing vectors are still designed adaptively based on the proposed active sensing scheme.

\subsubsection*{Random $\bm{w}^{(t)}$ with Perfect CSI}
Given the perfect CSI of the channels, we compute the downlink beamforming matrix $\bm{B}^{(t)}$ via \eqref{eq: fixed point euqation 1}-\eqref{eq: fixed point euqation 3} with randomly generated downlink RIS reflection coefficients $\bm{w}^{(t)}$.

\begin{figure}[t]
	\includegraphics[width=0.88\linewidth]{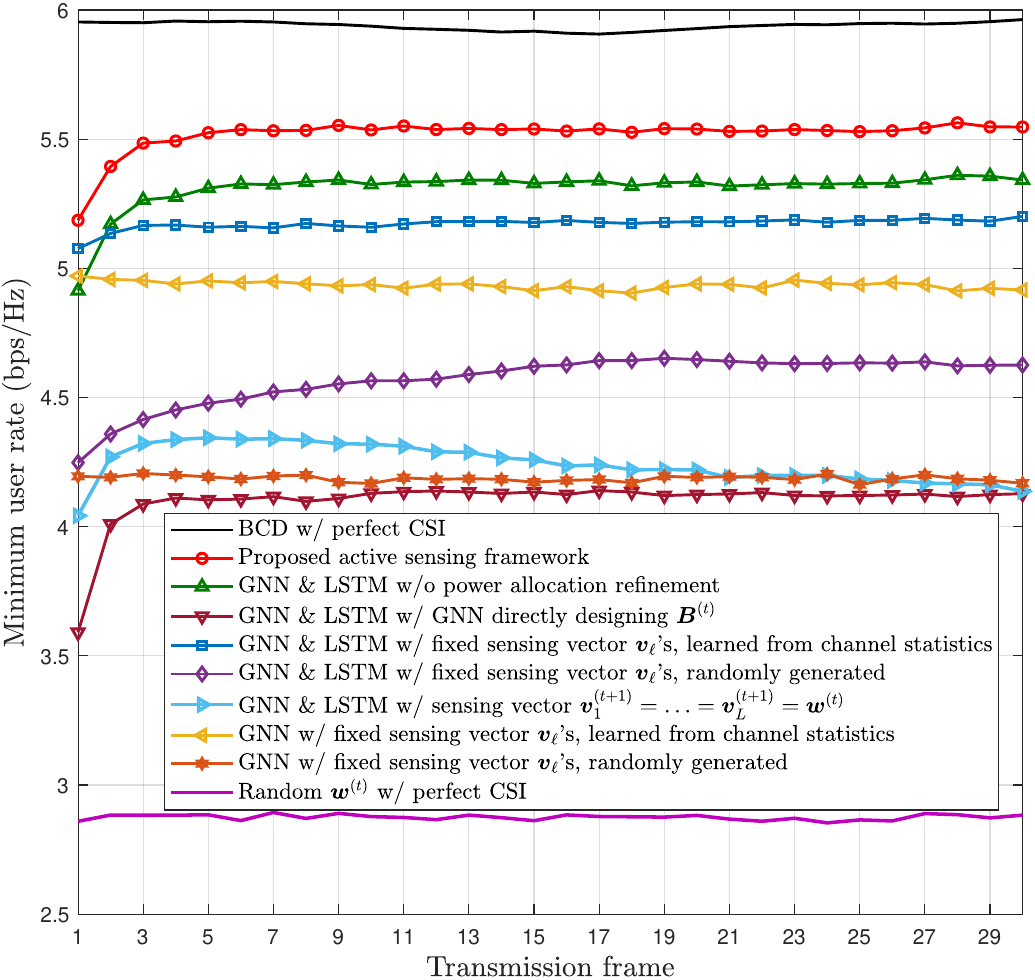}%
	\centering
	\caption{Average minimum user rate at the $1$-st block of each frame (i.e., $\min_k \{R_k^{(t,1)}\}$) for different beam tracking methods. Here, $L=3$, $M=8$, $N_r=100$, $K=3$, $N_G=8$, $\tau_w=10$, $P_u = 5\text{dBm}$, and $P_d = 10\text{dBm}$.} 
	\label{fig: ULnoise84_down90_rate_vs_frame}
\end{figure}

\subsection{Performance Evaluation}
Fig.~\ref{fig: ULnoise84_down90_rate_vs_frame} shows the downlink minimum user rate of different beam tracking methods in different transmission frames. The proposed active sensing framework initially achieves around $84\%$ of the BCD algorithm with perfect CSI, indicating efficient initial design based on the statistics of the UE mobility pattern. As the AP receives more pilots, the performance of the proposed approach improves quickly to $94\%$ of the performance of the benchmark with perfect CSI. This indicates that the proposed active sensing framework can efficiently exploit the temporal correlations of the time varying channel. Specifically, the beam tracking methods using both GNN and LSTM significantly outperform those benchmarks with only GNNs. This is because the LSTM-based methods are able to exploit the historical channel information to design the AP beamforming vectors and the downlink RIS reflection coefficients, thereby improving beam tracking performance over time. Among the beam tracking approaches with LSTM, the proposed framework performs the best, because it designs the RIS sensing vectors in an active manner. This shows the benefits of adaptively designing the RIS sensing vectors based on previously received pilots. 

For the benchmarks with fixed sensing schemes, Fig.~\ref{fig: ULnoise84_down90_rate_vs_frame} shows that the ones with learned sensing vectors outperform those with random sensing vectors. This shows the performance gain of designing the RIS sensing vectors based on channel statistics. Additionally, the GNN \& LSTM benchmark that uses the same RIS reflection coefficients for both pilot acquisition and data transmission stages performs less effectively as compared to the other GNN \& LSTM benchmarks that employ distinct RIS reflection coefficients for these two stages. This shows the importance of designing the RIS reflection coefficients in the channel sensing and data transmission stages separately. We also observe that the proposed approach significantly outperforms the GNN \& LSTM benchmark with GNN directly designing the beamforming matrix $\bm{B}^{(t)}$ based on pilots from the first $L$ sub-blocks of each frame. This advantage arises from leveraging the beamforming structure during the training stage, enabling more efficient training and better coordination between the downlink AP beamformers and the RIS reflection coefficients. Moreover, we remark that a $5\%$ performance improvement can be obtained from power allocation refinement.

As shown in Fig.~\ref{fig: ULnoise84_down90_rate_vs_frame}, the average minimum user rate obtained by iteratively optimizing the beamforming matrix $\bm{B}^{(t)}$ and $\bm{w}^{(t)}$ is approximately $6 \ \text{bps/Hz}$, compared to less than $3 \ \text{bps/Hz}$ when $\bm{w}^{(t)}$ is generated randomly. This demonstrates that the RIS plays a crucial role in beam tracking by significantly enhancing the communication channels between the AP and the mobile UEs. Furthermore, Fig.~\ref{fig: ULnoise84_down90_rate_vs_frame} shows that the proposed framework maintains its performance for transmission frames beyond $20$, despite being trained with only $20$ concatenated active sensing units corresponding to $20$ frames. We also observe empirically that the proposed framework continues to sustain its performance over hundreds of frames by reusing the active sensing unit. This robust performance beyond the training range is due to the ability of the LSTM cells to effectively capture the UE mobility patterns and to consistently update channel information based on the latest received pilots.

\begin{figure}[t]
	\includegraphics[width=0.75\linewidth]{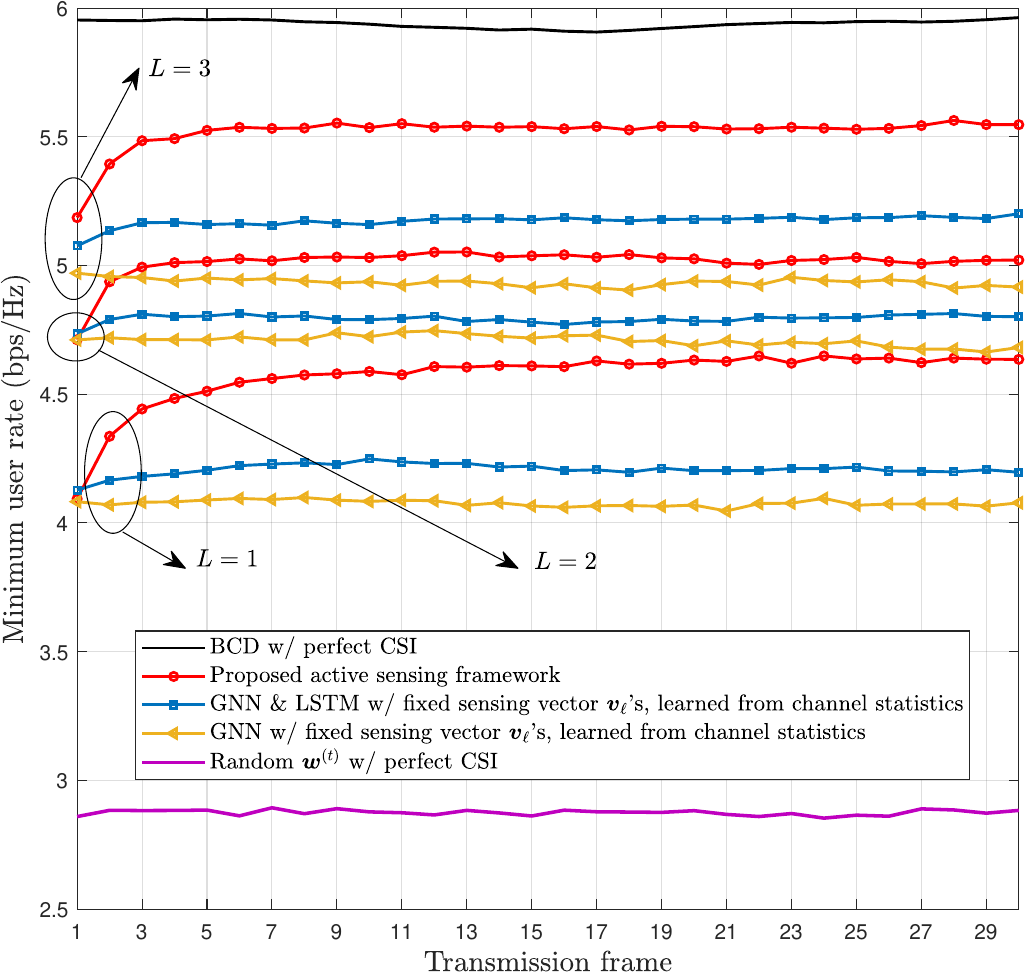}%
	\centering
	\caption{The impact of different numbers of sensing vectors $L$ within each frame on different beam tracking methods. Here, $M=8$, $N_r=100$, $K=3$, $N_G=8$, $\tau_w=10$, $P_u = 5\text{dBm}$, and $P_d = 10\text{dBm}$.} 
	\label{fig: ULnoise84_down90_different_L}
\end{figure}

\subsubsection*{Impact of Different $L$'s} 
Next, we evaluate the performance of the proposed approach for different numbers of sensing vectors $L$ within each frame. As can be seen from Fig.~\ref{fig: ULnoise84_down90_different_L}, the proposed active sensing framework always outperforms the benchmarks with fixed sensing strategies, because it leverages past observations to actively design the RIS sensing vectors. Furthermore, it is observed that LSTM-based methods (including the proposed approach) show faster convergence as the number of sensing sub-blocks $L$ increases. This is because a larger $L$ provides the LSTM with more observations in every frame, thereby facilitating more efficient capturing of temporal channel correlations.

\begin{figure}[t]
	\includegraphics[width=0.93\linewidth]{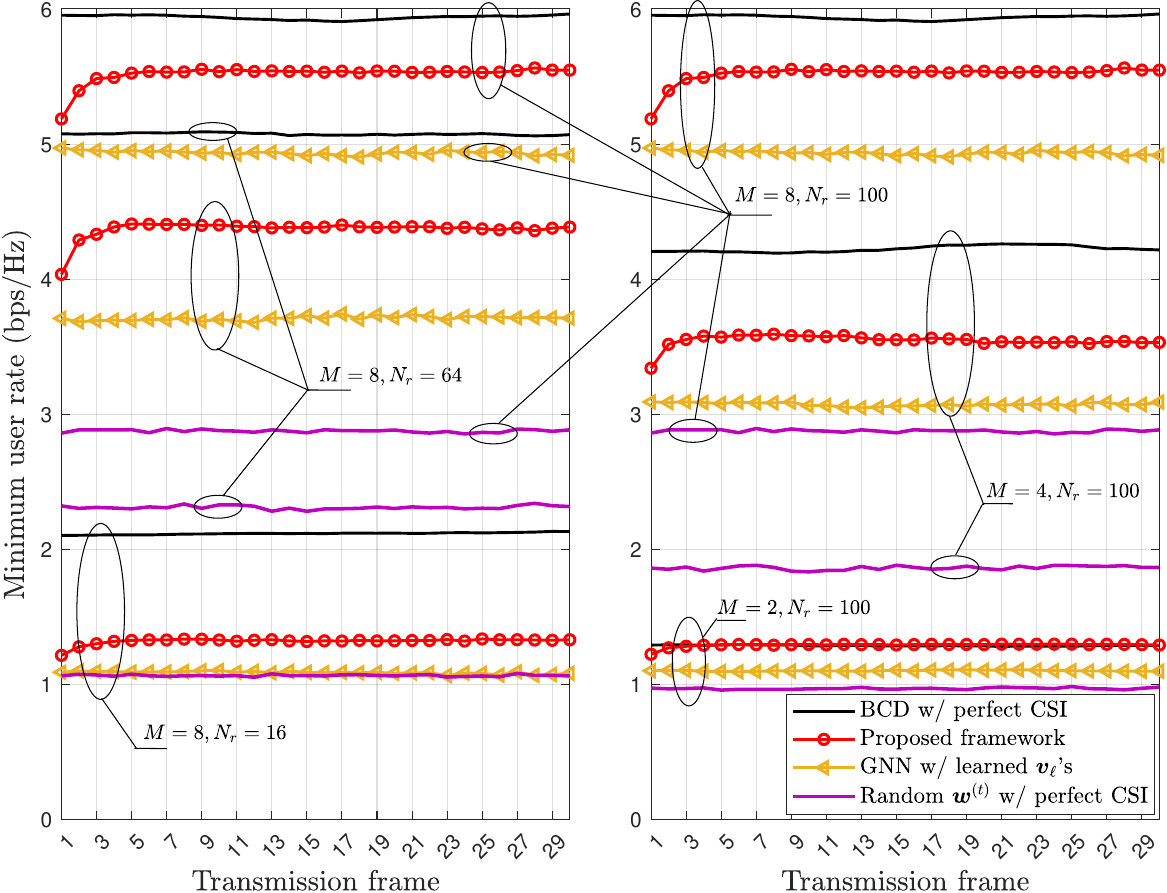}%
	\centering
	\caption{The impact of different numbers of AP antennas $M$ and RIS elements $N_r$ on the performance of the proposed approach. Here, $K=3$, $N_G=8$, $\tau_w=10$, $P_u = 5\text{dBm}$, and $P_d = 10\text{dBm}$.}
	\label{fig: ULnoise84_down90_different_M_Nr}
\end{figure}

\subsubsection*{Impact of Different $M$ and $N_r$} 
In this scenario, we fix the number of UEs at $K = 3$ and evaluate the performance of the proposed approach with varying numbers of AP antennas $M$ and RIS elements $N_r$. As shown in Fig.~\ref{fig: ULnoise84_down90_different_M_Nr}, the system performance (as measured by the minimum user rate) improves as the number of AP antennas and the number of RIS elements increase. Additionally, the proposed active sensing framework enhances tracking performance by up to $10\%$ across different configurations of $M$ and $N_r$. Moreover, the proposed approach consistently outperforms the GNN benchmark, which uses fixed sensing vectors learned from channel statistics.

\begin{figure}[t]
	\includegraphics[width=1\linewidth]{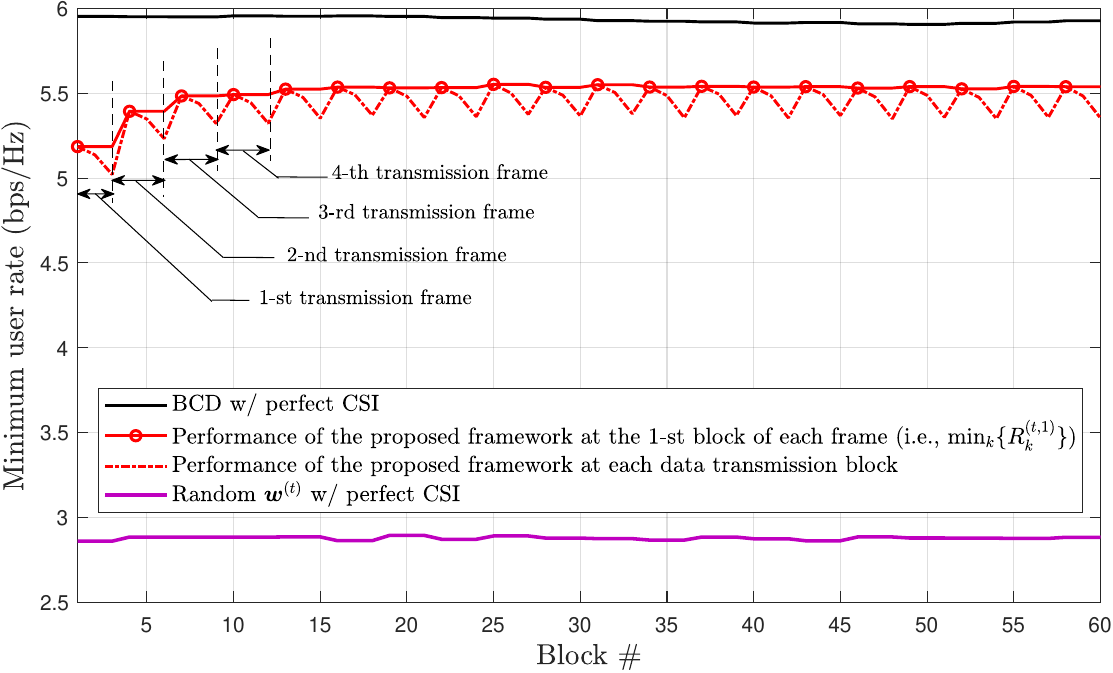}%
	\centering
	\caption{Average minimum user rate versus block number for the proposed active sensing framework. Here, $M=8$, $N_r=100$, $K=3$, $N_G=8$, $\tau_w=10$, $P_u = 5\text{dBm}$, and $P_d = 10\text{dBm}$.}
	\label{fig: fixed_beamforming}
\end{figure}

\subsubsection*{Effectiveness of Fixing Beamforming Strategy Across the Frame} 
Here, we illustrate the effect of using a fixed downlink beamforming strategy within each frame. In Fig.~\ref{fig: fixed_beamforming}, the dashed line shows the performance of the proposed framework in term of the minimum user data rate using fixed downlink beamforming vectors and RIS reflection coefficients, as the channels vary from the first block to the subsequent blocks within the frame. We see that the minimum data rate slightly decreases by about $3.5\%$ due to reusing the downlink beamforming vectors and RIS reflection coefficients for $N=3$ blocks within each data transmission stage, as the channels slowly evolve over time. In practice, the number of data transmission blocks per frame $N$ can be adjusted based on the environment's dynamics, such as the speed of the mobile UEs, to further optimize the balance between pilot overhead and performance.

\subsection{Scalability and Complexity Analysis}
In terms of computational complexity, it is important to note that the proposed learning-based approach is trained offline, so the main concern is inference complexity. The inference phase mainly involves matrix-vector multiplications, with the complexity scaling per LSTM cell per transmission frame is $\mathcal{O}(d^2 + df)$ where $d$ is the size of the hidden state and cell state, and $f$ is the dimension of the vectorized input $\bar{\bm{Y}}_{k}^{(t)}$. Given that $f$ is typically much smaller than $d$ and there are $K$ LSTM cells corresponding to $K$ mobile UEs, the total complexity for the LSTM cells per transmission frame is $\mathcal{O}(K d^2)$ after discarding the lower order terms. On the other hand, the complexity of the GNN is $\mathcal{O}(Kd\eta + (K + N_rL)\kappa + (K+2)^2 \varsigma D)$, where $D$, $\eta$, $\kappa$, and $\varsigma$ denote the number of updating layers in the GNN, the dimension of the input layer, the dimension of the output layer, and the computational complexity of the aggregation and combination operations, respectively. Therefore, the total computational complexity for the proposed active sensing unit during the inference stage is $\mathcal{O}(K d^2 + Kd\eta + (K + N_rL)\kappa + (K+2)^2 \varsigma D)$. It is important to note that the inference process is highly parallelizable using modern graphic process units (GPUs), so that in practice the proposed approach can be executed very efficiently.

We note that there are several ways to further reduce the computational complexity of the proposed approach in practical implementations, particularly for large-scale networks with many UEs $K$ and RIS elements $N_r$. First, the $K$ copies of the LSTM cells can be executed in parallel, making the actual complexity equivalent to the runtime of a single LSTM cell. Second, as shown in Fig.~\ref{fig: ULnoise84_down90_different_L}, the proposed active sensing framework can achieve satisfactory performance with very few sensing vectors. Thus, only a few RIS sensing vectors need to be designed, thereby conserving computational resources. Finally, while we opt for relatively large neural networks in the simulations to characterize ultimate performance, we find that reducing the size of the neural network does not significantly impact performance.

\subsection{Interpretation of Learned Solutions}
In this section, we use the array response of the RIS and the SINR map as a means to illustrate qualitatively and quantitatively the correctness of the solutions learned by the proposed framework, respectively. In the pilot stage (i.e., $0$-th block) of the $t$-th frame, the array response of the RIS sensing vector $\bm{v}_{\ell}^{(t)}$ at the coordinates $[x,y]$ is given as \cite{jiang2021learning}:
\begin{equation} \label{RIS sensing array response}
    \begin{aligned}
        & r_{\text{RIS}}\left(\phi_{2,1},\dots, \phi_{2,N_G}, \theta_{3,[x,y]}^{(t,0)}, \phi_{3,[x,y]}^{(t,0)}, \bm{v}_{\ell}^{(t)}\right) \\
        & \!=\! \left|\left(\sum_{i=1}^{N_G} \bm{a}_{\text{RIS}}\left(\theta_{2,i}, \phi_{2,i}\right)\!\right)^{\sf{H}} \!\!\!\!\operatorname{diag}\left(\bm{v}_{\ell}^{(t)}\right) \bm{a}_{\text{RIS}}\!\left(\theta_{3,[x,y]}^{(t,0)}, \phi_{3,[x,y]}^{(t,0)}\right)\right| ,
    \end{aligned}
\end{equation}
where $\ell = 1, \dots, L$ and $\theta_{2,i} = 0$, $i = 1,\dots,N_G$. Here, $\theta_{3,[x,y]}^{(t,n)}$ and $\phi_{3,[x,y]}^{(t,n)}$ denotes the elevation and azimuth AoA from the coordinates $[x,y]$ to the RIS. When the coordinates $[x,y]$ coincide with the actual location of the mobile UE $k$, $\theta_{3,[x,y]}^{(t,n)}$ and $\phi_{3,[x,y]}^{(t,n)}$ equivalents to $\theta_{3,k}^{(t,n)}$ and $\phi_{3,k}^{(t,n)}$ as in \eqref{RIS steering vector}. Similarly, the array response of the downlink RIS reflection coefficients $\bm{w}^{(t)}$ in the $n$-th block of the $t$-th frame can be computed as $r_{\text{RIS}}(\phi_{2,1},\dots, \phi_{2,N_G}, \theta_{3,[x,y]}^{(t,n)}, \phi_{3,[x,y]}^{(t,n)}, \bm{w}^{(t)})$. In addition, the array response of the downlink beamforming vector $\bm{b}_k^{(t)}$ can be expressed as:
\begin{equation}
    r_{\text{AP},k}(\phi,\bm{b}_k^{(t)}) = \left|\bm{a}_{\text{AP}}(\phi)^{\sf{H}} \bm{b}_k^{(t)}\right| , k = 1, \dots, K ,
\end{equation}
where $\phi \sim \mathcal{U}(0,2\pi)$ is the effective AoA. We can further compute the SINR using the downlink beamformers $\{\bm{b}^{(t)}_k\}_{k=1}^K$ and RIS reflection coefficients $\bm{w}^{(t)}$ as follows:
\begin{equation}
    \frac{\left|\left(\bm{A}^{(t,n)}_{[x,y]} \widetilde{\bm{w}}^{(t)}\right)^{\sf{H}} \bm{b}_k^{(t)}\right|^2}{\sum_{j=1, j \neq k}^K \left|\left(\bm{A}^{(t,n)}_{[x,y]} \widetilde{\bm{w}}^{(t)}\right)^{\sf{H}} \bm{b}_j^{(t)}\right|^2+\sigma_d^2} , \ k = 1,\dots,K ,
\end{equation}
where $\bm{A}^{(t,n)}_{[x,y]}$ is the combined channel between the AP and the coordinates $[x,y]$ in the $n$-th block of the $t$-th frame. When the coordinates $[x,y]$ coincide with the actual location of the mobile UE $k$, the combined channel $\bm{A}^{(t,n)}_{[x,y]}$ equivalents to $\bm{A}^{(t,n)}_{k}$ as in \eqref{eq:DL signal}.

\begin{figure*}[t]
    \centering
    \subfigure[Array responses of the downlink RIS reflection coefficients $\bm{w}^{(t)}$ obtained by the proposed framework.]{
        \begin{minipage}[t]{1\textwidth}
            \centering
            \includegraphics[width=0.945\linewidth]{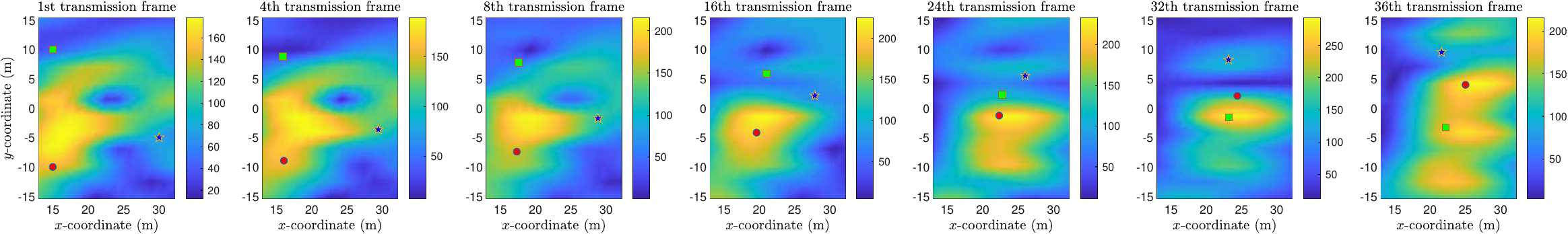}
            \label{fig: AR downlink RIS}
        \end{minipage}%
    }
    \subfigure[Array responses of the downlink beamforming vectors $\bm{b}^{(t)}_k$ obtained by the proposed framework.]{
        \begin{minipage}[t]{0.98\textwidth}
            \centering
            \includegraphics[width=0.945\linewidth]{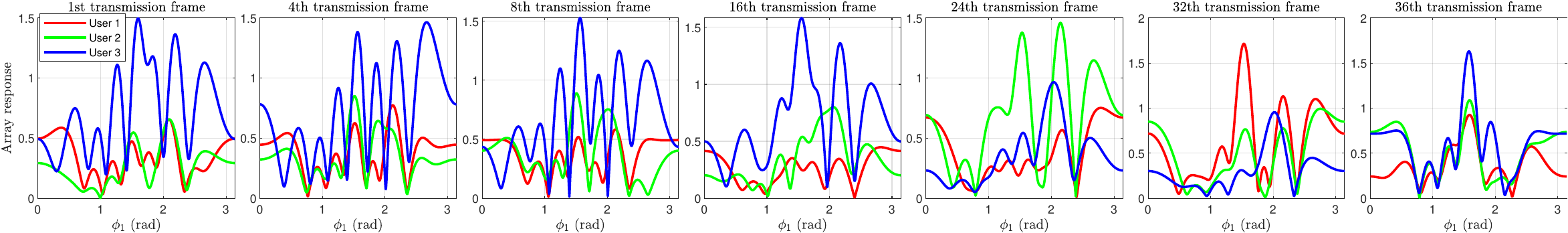}
            \label{fig: AR beamformers}
        \end{minipage}
    }
\vspace{-2mm}
\caption{Array responses of the downlink RIS reflection coefficients and beamformers obtained by the proposed framework in different frames. The proposed framework is tested on a randomly generated trajectory. Here, $L=3$, $M=8$, $N_r=100$, $K=3$, $N_G = 8$, $\tau_w = 10$, $P_u = 5\text{dBm}$, and $P_d = 10\text{dBm}$.}
\label{fig: DL array response - DLnoise90}
\end{figure*}

\begin{figure*}[!h]
    \centering
    \vspace{-2mm}
    \subfigure[SINR (dB) maps of the UE $1$.]{
        \begin{minipage}[t]{1\textwidth}
            \centering
            \includegraphics[width=0.945\linewidth]{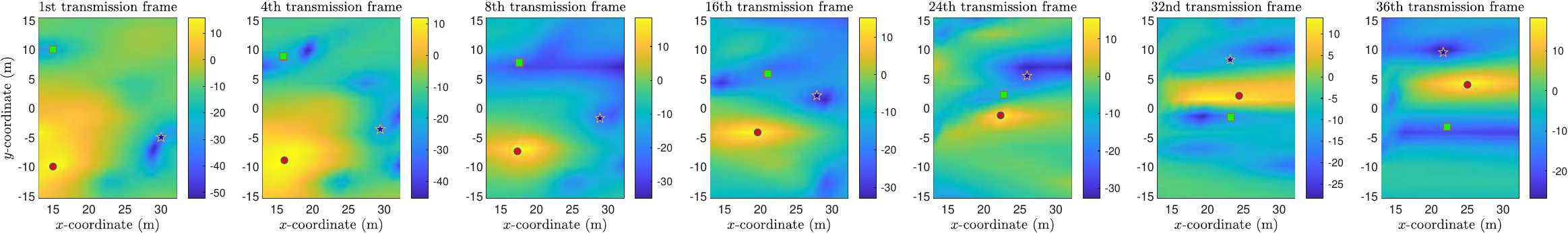}
            \label{fig: SINR 1}
        \end{minipage}%
    }
    \vspace{-2mm}
    \subfigure[SINR (dB) maps of the UE $2$.]{
        \begin{minipage}[t]{1\textwidth}
            \centering
            \includegraphics[width=0.945\linewidth]{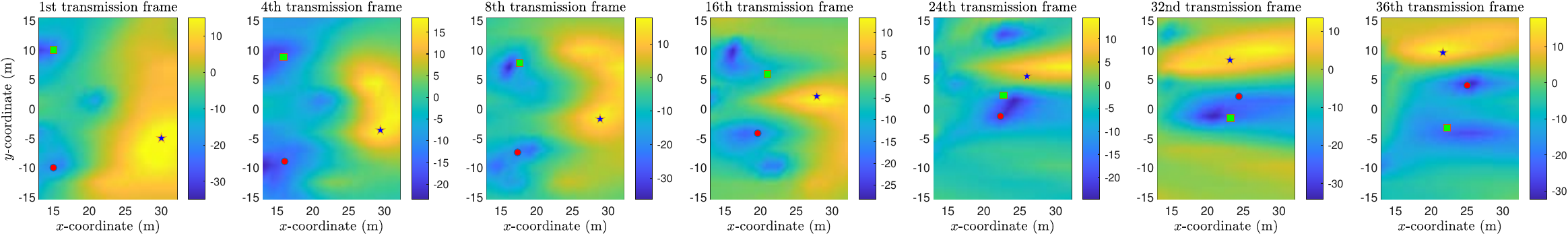}
            \label{fig: SINR 2}
        \end{minipage}
    }
    \vspace{-3mm}
    \subfigure[SINR (dB) maps of the UE $3$.]{
        \begin{minipage}[t]{1\textwidth}
            \centering
            \includegraphics[width=0.945\linewidth]{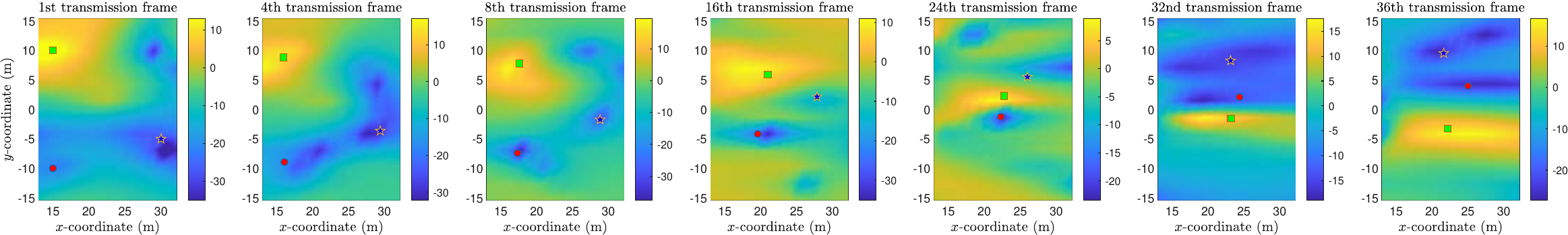}
            \label{fig: SINR 3}
        \end{minipage}
    }
    \vspace{-1mm}
\caption{SINR (dB) maps around the positions of mobile UEs under the proposed active sensing framework in different frames. The proposed framework is tested on a randomly generated trajectory. Here, $L=3$, $M=8$, $N_r=100$, $K=3$, $N_G = 8$, $\tau_w = 10$, $P_u = 5\text{dBm}$, and $P_d = 10\text{dBm}$.}
\label{fig: DL SINR - DLnoise90}
\end{figure*}

\subsubsection*{Interpretation of Downlink Beam Alignment Strategy}
In Fig.~\ref{fig: DL array response - DLnoise90}(a), the array response of the downlink RIS reflection coefficients $\bm{w}^{(t)}$ shows that the proposed framework effectively tracks the UEs by progressively reflecting increasingly focused downlink beams via the RIS towards the actual positions of the mobile UEs. It is observed that the peaks corresponding to the first two UEs are sometimes weaker than the third UE, but we can see from Fig.~\ref{fig: DL array response - DLnoise90}(b) that these weaker UEs are compensated by stronger array response of the beamformers at the AP. Therefore, the proposed framework indeed learns to jointly optimize the downlink RIS reflection coefficients and the beamforming matrix to ensure fairness in maximizing the minimum rate across the three mobile UEs, while exploiting channel correlations to design improved RIS reflection coefficients and beamformers over time. 

\subsubsection*{Interpretation of Learned Solutions via SINR Map}
we further plot the downlink SINR map for each mobile UE to show the overall beam tracking effectiveness of the proposed framework. As shown in Fig.~\ref{fig: DL SINR - DLnoise90}, the SINR map broadly covers areas around the UEs initially, but progressively narrows to focus on actual positions of all three mobile UEs. This progressive focusing indicates that proposed framework is able to design sensible downlink beamformers $\bm{b}_k^{(t)}$ and data transmission RIS reflection coefficients $\bm{w}^{(t)}$ by utilizing both temporal and spatial information from the received pilots. 

\subsubsection*{Interpretation of the Proposed Active Sensing Strategy}
\begin{figure*}[t]
\centering
\subfigure[$\bm{v}^{(t)}_1$ actively designed by the proposed framework.]{
\includegraphics[width=.46\textwidth]{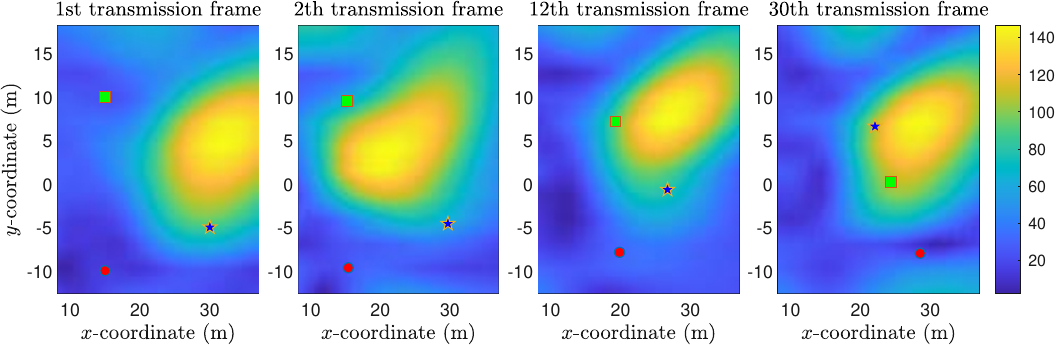}
} 
\hspace{3mm}
\subfigure[Fixed $\bm{v}_1$, learned from channel statistics.]{
\includegraphics[width=.46\textwidth]{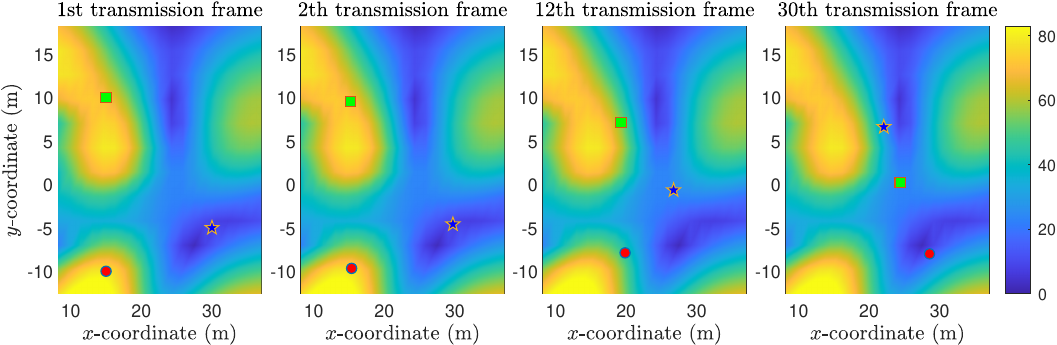}
}
\vspace{-2mm}

\subfigure[$\bm{v}^{(t)}_2$ actively designed by the proposed framework.]{
\includegraphics[width=.46\textwidth]{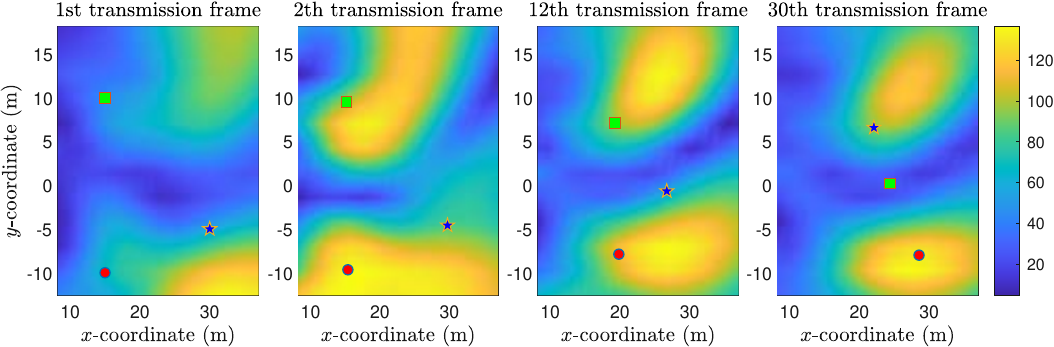}
} 
\hspace{3mm}
\subfigure[Fixed $\bm{v}_2$, learned from channel statistics.]{
\includegraphics[width=.46\textwidth]{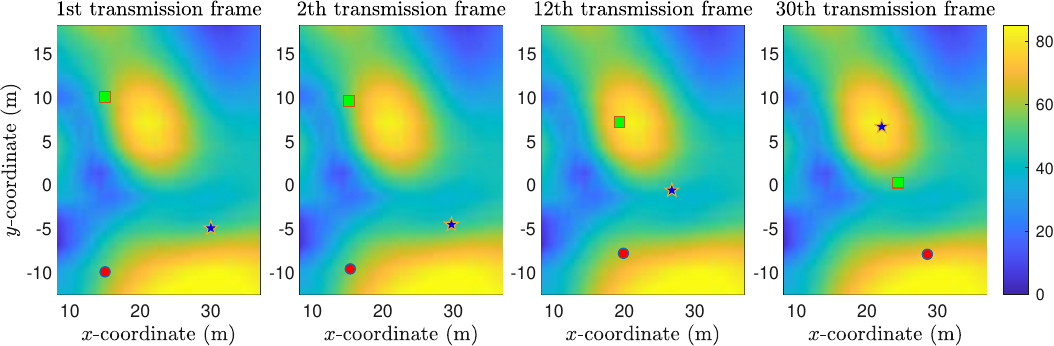}
}
\vspace{-2mm}

\subfigure[$\bm{v}^{(t)}_3$ actively designed by the proposed framework.]{
\includegraphics[width=.46\textwidth]{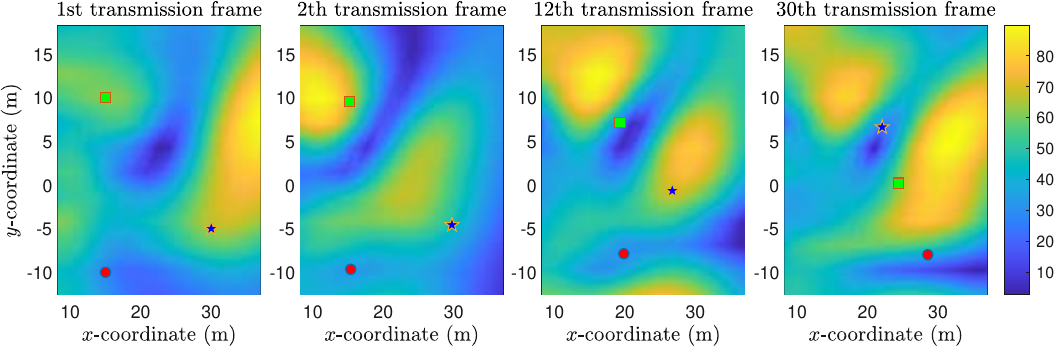}
} 
\hspace{3mm}
\subfigure[Fixed $\bm{v}_3$, learned from channel statistics.]{
\includegraphics[width=.46\textwidth]{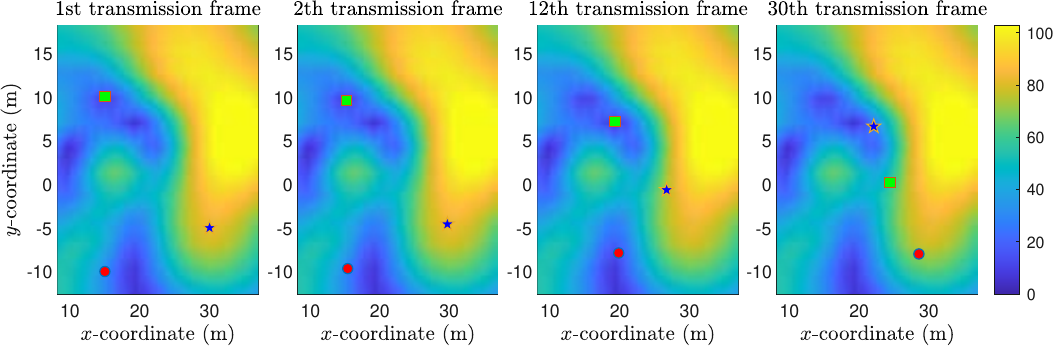}
}
\vspace{-1mm}
\caption{The array responses of the RIS sensing vectors obtained by the proposed framework (left) and by the GNN \& LSTM Without Active Sensing benchmark (right) in the pilot stage (the 0-th block) across different frames. Here, $L=3$, $M=8$, $N_r=100$, $K=3$, $N_G = 8$, $\tau_w = 10$, $P_u = 5\text{dBm}$, and $P_d = 10\text{dBm}$.}
\label{fig: sensing vector comparison}
\end{figure*}

As illustrated in Fig.~\ref{fig: sensing vector comparison}, the proposed framework designs $L=3$ RIS sensing vectors cooperatively. For example, in the pilot stage of the $30$-th frame, one UE (represented by the red dot) is not targeted by the first RIS sensing vector $\bm{v}_1^{(t)}$, but is focused on by $\bm{v}_2^{(t)}$. Meanwhile, the weaker array response of $\bm{v}_2^{(t)}$ towards the green UE and $\bm{v}_3^{(t)}$ towards the blue UE are compensated by stronger responses from the other two sensing vectors, respectively. This cooperative sensing strategy is not seen in the initial tracking frame, but is progressively developed over time by leveraging historical channel observations. As a result, the proposed framework actively narrows the sensing range and enhances the pilot SNR in the uplink CSI acquisition stages, leading to more informative neural network inputs and improved beam tracking performance. In contrast, the GNN \& LSTM Without Active Sensing benchmark employs a non-adaptive approach by using channel statistics to train fixed RIS sensing vectors, which cover a larger area for CSI acquisition. However, without adaptive focusing, this approach risks losing track of certain UEs; for instance, the green UE is not as well focused on in the $30$-th frame as compared to the proposed active sensing strategy. The broader coverage would in general result in lower uplink SNR during CSI acquisition and, subsequently, higher pilot overhead.

\section{Conclusion} \label{conclusion}
This paper proposes an active sensing based beam tracking approach for RIS-assisted multiuser mobile communication systems. The proposed approach can adaptively design the AP beamformers and the RIS reflection coefficients for downlink data transmission, as well as the RIS reflection coefficients for uplink channel sensing. This is accomplished by utilizing an LSTM-based RNN to summarize received pilot information into state vectors, and a GNN to optimize the mapping from the state vectors to the beam tracking matrices. Numerical results demonstrate that the proposed active sensing framework produces interpretable solutions and achieves significantly higher data rates than benchmark data-driven methods with non-adaptive sensing.

\bibliography{bibliography}
\bibliographystyle{IEEEtran}

\begin{IEEEbiography}[{\includegraphics[width=1.2in,height=1.25in,clip,keepaspectratio]{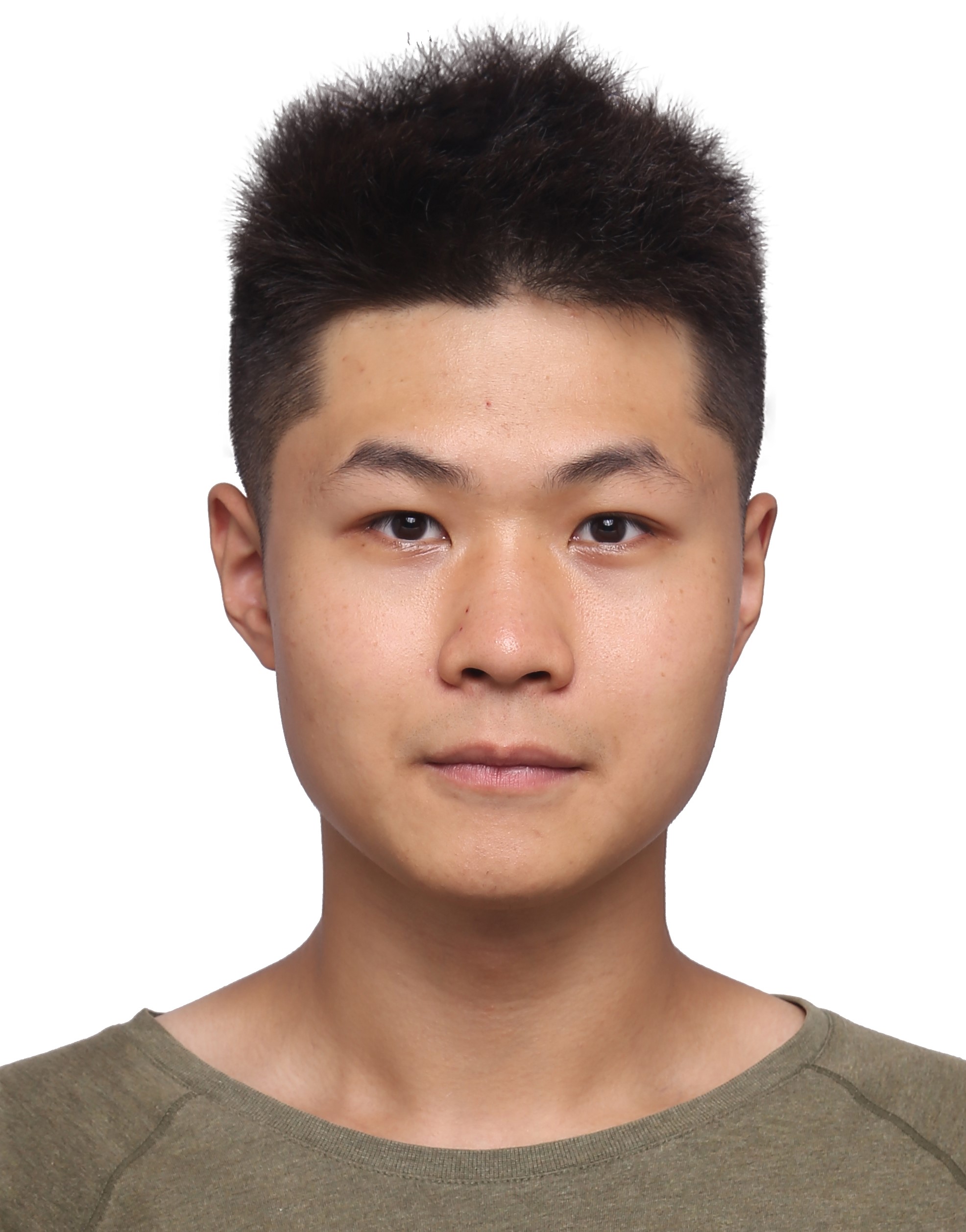}}]{Han Han} (Member, IEEE)  received the B.E. degree in electronic information engineering from the University of Electronic Science and Technology of China, Chengdu, China, in 2021, and the M.A.Sc. degree in electrical and computer engineering from the University of Toronto, Toronto, ON, Canada, in 2024. He is now with Marvell Semiconductor Canada, Inc., Toronto, ON, Canada. His main research interests include wireless communications, machine learning, and digital signal processing.
\end{IEEEbiography}

\begin{IEEEbiography}[{\includegraphics[width=1.2in,height=1.25in,clip,keepaspectratio]{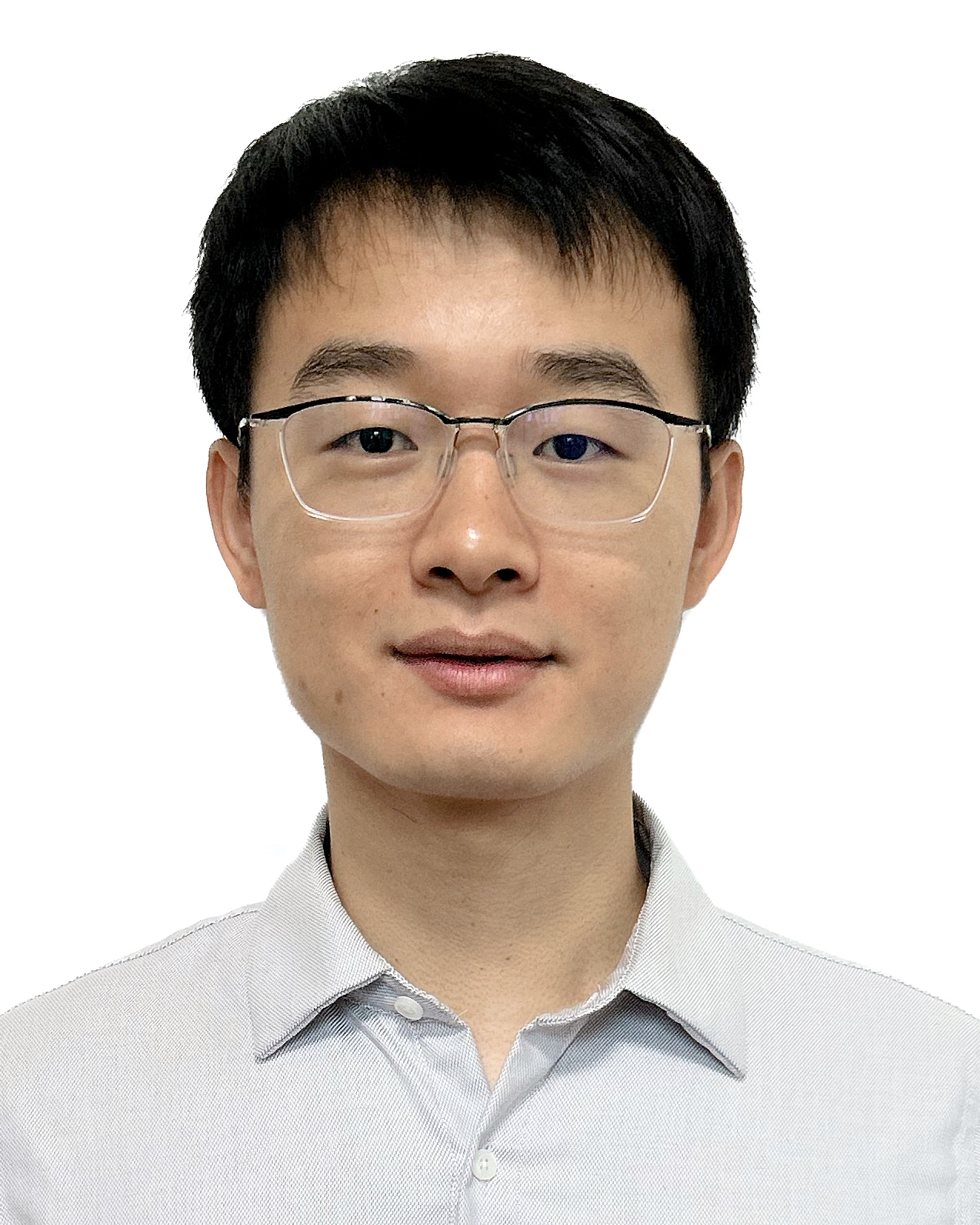}}]{Tao Jiang} (Member, IEEE)  received the B.E. degree in telecommunications engineering from Xidian University, Xi'an, China, in 2017, the M.E. degree in computer science from ShanghaiTech University, Shanghai, China, in 2020, and the Ph.D. degree in electrical and computer engineering from University of Toronto, Toronto, ON, Canada, in 2024. He is now with Qualcomm Technologies, Inc., Santa Clara, CA, USA. His main research interests include wireless communications, machine learning and optimization.
\end{IEEEbiography}

\begin{IEEEbiography}[{\includegraphics[width=1.2in,height=1.25in,clip,keepaspectratio]{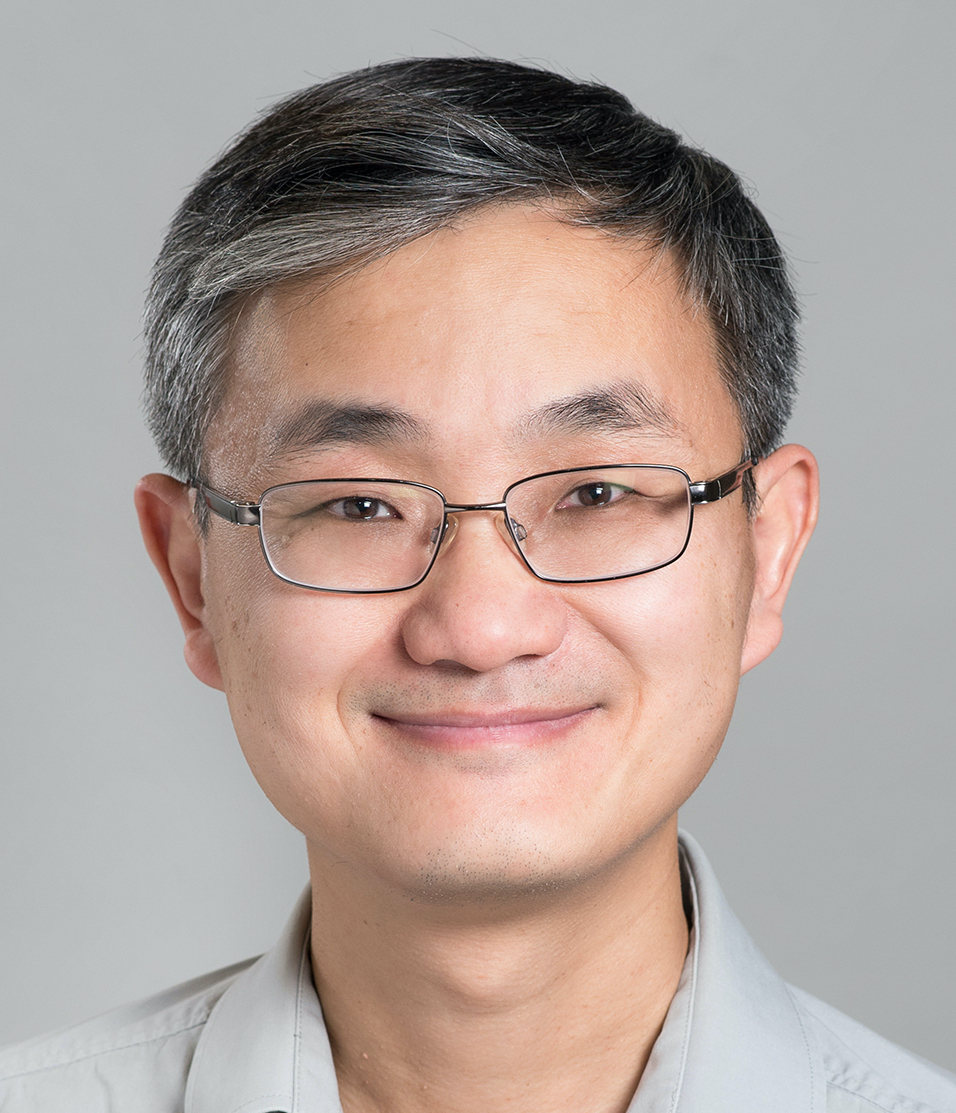}}]{Wei Yu} (Fellow, IEEE)  received the B.A.Sc. degree in computer engineering and mathematics from the University of Waterloo, Waterloo, ON, Canada, and the M.S. and Ph.D. degrees in electrical engineering from Stanford University, Stanford, CA, USA. He is currently a Professor in the Electrical and Computer Engineering Department at the University of Toronto, Toronto, ON, Canada, where he holds a Canada Research Chair (Tier 1) in Information Theory and Wireless Communications. He is a Fellow of the Canadian Academy of Engineering and a member of the College of New Scholars, Artists, and Scientists of the Royal Society of Canada. Prof. Wei Yu was the President of the IEEE Information Theory Society in 2021 and served on its Board of Governors in 2015-2023. He served as the Chair of the Signal Processing for Communications and Networking Technical Committee of the IEEE Signal Processing Society in 2017-2018. He was an IEEE Communications Society Distinguished Lecturer in 2015-2016. He served as an Area Editor of the IEEE Transactions on Wireless Communications, as an Associate Editor for IEEE Transactions on Information Theory, and as an Editor for the IEEE Transactions on Communications and IEEE Transactions on Wireless Communications. Prof. Wei Yu received the IEEE Communications Society and Information Theory Society Joint Paper Award in 2024, the IEEE Signal Processing Society Best Paper Award in 2021, 2017, and 2008, the IEEE Marconi Prize Paper Award in Wireless Communications in 2019, the IEEE Communications Society Award for Advances in Communication in 2019, the Journal of Communications and Networks Best Paper Award in 2017, the IEEE Communications Society Best Tutorial Paper Award in 2015, and the Steacie Memorial Fellowship in 2015. He is a Clarivate Highly Cited Researcher.
\end{IEEEbiography}

\end{document}